\documentclass[12pt]{article}

\usepackage{newtxtext,newtxmath}

\usepackage{graphicx}

\usepackage[letterpaper,margin=1in]{geometry}

\renewenvironment{abstract}
	{\quotation}
	{\endquotation}

\date{}

\makeatletter
\renewcommand{\fnum@figure}{\textbf{Figure \thefigure}}
\renewcommand{\fnum@table}{\textbf{Table \thetable}}
\makeatother

\usepackage{scicite}

\usepackage{url}

\def\scititle{
	\textit{In situ} Evidence of 5-minute Oscillations from Parker Solar Probe
}
\title{\bfseries \boldmath \scititle}

\author{
	Zesen~Huang$^{1\ast\dagger}$,
	Marco~Velli$^{1\ast}$,
	Olga Panasenco$^{2}$,
	Richard J. Morton$^{3}$,
	Chen~Shi$^{4}$,\\
	Yeimy J. Rivera$^{5}$,
    Benjamin Chandran$^{6}$,
    Samuel T. Badman$^{5}$,
	Yuliang Ding$^{1}$,\\
	Nour Raouafi$^{7}$,
	Stuart D. Bale$^{8,9}$,
	Michael Stevens$^{5}$,\\
    Tamar Ervin$^{8,9}$,
	Chuanpeng Hou$^{10}$,
	Kristopher G. Klein$^{11}$,\\
	Orlando Romeo$^{8}$,
	Jia Huang$^{8}$,
	Mingzhe Liu$^{8}$,\\
	Davin E Larson$^{8}$
	Marc Pulupa$^{8}$,
	Roberto Livi$^{8}$,
	Federico Fraschetti$^{5}$
	\and
	\small$^{1}$Department of Earth, Planetary, and Space Sciences, University of California, Los Angeles; Los Angeles, 90049, USA\and
	\small$^{2}$Advanced Heliophysics, Pasadena, CA 91106, USA \and
	\small$^{3}$School of Engineering, Physics \& Mathematics, Northumbria University, Newcastle upon Tyne, NE2 3RA, UK \and
	\small$^{4}$Department of Physics, Auburn University; 380 Duncan Dr., Auburn, AL 36849, USA \and
	\small$^{5}$Center for Astrophysics, Harvard and Smithsonian, Cambridge, MA 02138, USA. \and
	\small$^{6}$Department of Physics \& Astronomy, University of New Hampshire, Durham, NH, USA \and
	\small$^{7}$Johns Hopkins University Applied Physics Laboratory, Laurel, MD, USA \and
	\small$^{8}$Space Sciences Laboratory, University of California, Berkeley, CA 94720, USA. \and
	\small$^{9}$Department of Physics, University of California, Berkeley, CA 94720, USA. \and
	\small$^{10}$University of Potsdam, Karl-Liebknecht-Str. 24-25, Haus 28, 2.109, 14476 Potsdam, Germany \and
	\small$^{11}$Department of Planetary Sciences, University of Arizona, Tucson, AZ, USA \and
	\small$^\ast$Corresponding author. Email: zesenhuang@g.ucla.edu, mvelli@ucla.edu\and
	\small$^\dagger$These authors contributed equally to this work.
}

\begin{document} 

\maketitle

\begin{abstract} \bfseries \boldmath
The Sun’s surface vibrates in characteristic 5-minute oscillations, known as \textit{p}-modes, generated by sound waves trapped within the convection zone. Although these oscillations have long been hypothesized to reach into the solar wind, direct in situ evidence has remained elusive, even during previous close encounters by Parker Solar Probe (PSP). Here, we present the first definitive in situ detection of 5-minute oscillations in the upper solar corona, based on observations from PSP’s three closest perihelia. In two events at 9.9 solar radii ($R_\odot$), we identify statistically significant ($\sim$6$\sigma$) 3.1–3.2 mHz peaks in the magnetic field power spectrum, each appearing as a large-amplitude, spherically polarized Alfvénic wave train lasting approximately 35 minutes. These results demonstrate that global solar oscillations can reach and potentially influence the solar wind.
\end{abstract}
\clearpage

\noindent
The 5-minute oscillations were first identified in data acquired from the solar telescope at Mt. Wilson Observatory during the summers of 1960 and 1961 \cite{leighton_velocity_1962,noyes_velocity_1963,simon_velocity_1964}. These measurements captured the line-of-sight velocities of solar photospheric plasma, revealing oscillations primarily composed of locally in-phase wave packets with periods of $296 \pm 3$ seconds. These wave packets are typically short-lived, with a characteristic lifetime of about 30 minutes and remains in-phase over a typical range of 0.5 arcminute on the solar disk \cite{musman_vertical_1970}. The origin of the 5-minute oscillations was resolved in the early 1970s by \cite{ulrich_five-minute_1970} and \cite{leibacher_new_1971}, who demonstrated that sound waves generated by photospheric granulation are trapped within the convection zone, forming a resonance chamber. These resonances, now known as \textit{p}-modes in helioseismology, collectively produce the 5-minute oscillations \cite{deubner_helioseismology_1984,bahcall_solar_1988}.


The 5-minute oscillations are primarily compressible (acoustic) at the photosphere, whereas the dominant oscillations in the fast solar wind are Alfvén waves \cite{rivera_situ_2024,velli_propagation_1993,belcher_alfvenic_1971,heinemann_non-wkb_1980,hollweg_alfven_1973,morton_basal_2019,morton_alfvenic_2023,telloni_energy_2023,shi_acceleration_2022}, which are inherently incompressible. It has been proposed that sound waves in the photosphere can convert into Alfvén waves in the lower corona through a process known as double-mode conversion. In this scenario, sound waves first linearly convert into fast magneto-sonic modes at the magnetic canopy in the chromosphere, which then couple with Alfvén modes at higher altitudes \cite{cally_alfven_2012,cally_benchmarking_2011,hansen_benchmarking_2012,huang_conservation_2022,miriyala_coronal_2025}. This mechanism may transfer up to 30\% of the photospheric p-mode energy flux into the lower corona as Alfvén waves, potentially heating the solar corona and powering the solar wind. Although 5-minute oscillations have long been observed to reach into the chromosphere and solar corona \cite{pontieu_how_2005,didkovsky_observations_2011,guglielmi_five-minute_2015,calabro_oscillatory_2013,morton_origins_2025-1}, evidence for such oscillations in the solar wind has remained inconclusive and controversial \cite{thomson_propagation_1995,chashei_five-minute_1999,guglielmi_five-minute_2015}.

A direct explanation for the lack of evidence for 5-minute oscillations in the solar wind is the nonlinear evolution of Alfv\'en waves. At 1 AU, Alfv\'en wave energy is typically distributed evenly across logarithmic frequencies, forming the well-known $1/f$ range \cite{bruno_solar_2013}. As the heliocentric distance decreases, this energy gradually shifts to higher frequencies \cite{bruno_coordinated_2009,huang_two_2025}. Recent Parker Solar Probe (PSP) \cite{fox_solar_2016} observations show that the $1/f$ range is usually absent near the Alfv\'en (critical) surface—the upper boundary of the solar corona \cite{kasper_parker_2021}. Instead, Alfv\'en wave energy becomes concentrated around a 2-minute period \cite{huang_new_2023,davis_evolution_2023,huang_dominance_2024}. Due to the short propagation time of Alfv\'en waves from the lower corona, it has been proposed that observations near the Alfv\'en surface directly reflect the Alfv\'en wave energy spectrum in the lower corona \cite{huang_dominance_2024,huang_two_2025}. As expected, these findings are consistent with remote sensing observations from the Solar Dynamics Observatory (SDO) within 1.2 $R_\odot$ \cite{morton_alfvenic_2023,morton_basal_2019} and DKIST Cryo-NIRSP \cite{morton_origins_2025-1}, which show two distinct peaks corresponding to 5-minute and 2-minute oscillations. Thus, the unprecedented \textit{in situ} observations of the upper solar corona from PSP's final orbits may provide direct evidence for 5-minute oscillations that survive the solar corona and reach the primordial solar wind.

In this study, we analyze all available PSP observations within $\pm$7 days of perihelion for encounters E1 to E24 (where E\# denotes each PSP solar encounter) and identify two prominent 5-minute oscillation events from two of the three final-orbit encounters: one on 2024-12-24 (E22) and one on 2025-06-19 (E24). Both events occurred near the mission-wide perihelion at approximately 9.9 solar radii ($R_{\odot}$) and exhibit a distinct peak around 3.2 mHz (3.2 mHz and 3.1 mHz, respectively) in the rectified trace magnetic field power spectral density (PSD; see \cite{methods} for definition), with significances of 5.8$\sigma$ and 6.2$\sigma$, respectively. Given the strong similarities between these events, we focus on the first event as a representative example in the main text, with details of the second event provided in the supplementary materials.

\subsection*{Event Overview}
PSP reached its first mission-wide closest perihelion on 2024-12-24 at 11:53 UTC (Coordinated Universal Time). The interval from 09:50 to 12:05 encompasses the 5-minute oscillation event, as shown in Fig.~\ref{fig:event-p1} and Fig.~\ref{fig:event-p2}. The spacecraft entered from the top left of Fig.~\ref{fig:event-p1}\textbf{(a)}, attaining a perihelion of 9.86~$R_{\odot}$. Throughout this interval, highlighted by the pink shaded region in Fig.~\ref{fig:event-p2}\textbf{(a)} to \textbf{(f)}, plasma conditions remained quasi-static, with proton number densities relatively constant except for a brief spike near 11:00 (panel \textbf{(e)}). The solar wind was deeply sub-Alfv\'enic (see the red line in Fig.~\ref{fig:event-p1}\textbf{(a)} and Fig.~\ref{fig:event-p2}\textbf{(c)} and \textbf{(f)}), with Alfv\'en Mach number $M_A = V_r/V_A \lesssim 0.5$, where $V_r$ is the proton radial bulk speed and $V_A$ is the local Alfv\'en speed. This indicates that PSP was directly sampling the solar corona \cite{kasper_parker_2021}. Additionally, the magnetic field magnitude remained nearly constant (black line, panel \textbf{(a)}), the cross helicity $\sigma_c$ was close to unity (blue line, panel \textbf{(d)}), and the plasma $\beta$ was very low (black line, panel \textbf{(d)}), suggesting that the fluctuations were predominantly outward-propagating Alfv\'en waves. (For plasma parameter definitions, see \cite{methods})

To enable a fair comparison across frequencies in logarithmic space, for any given PSD $P(f)$, we also present the rectified PSD $P(f)\cdot f$ to visualize the energy distribution. This is because $P(f)\,\mathrm{d}f = P(f)\cdot f\,\mathrm{d}\ln f$ \cite{huang_dominance_2024,huang_two_2025}. The trace magnetic PSD computed from Fast Fourier Transform ($PSD_{FFT}$) is shown as the blue line in Fig.~\ref{fig:event-p1}\textbf{(b)}. The rectified wavelet power spectrogram $S(f,t)\cdot f$ is shown in Fig.~\ref{fig:event-p2}\textbf{(b)}, visualizing the time-frequency (period) distribution of fluctuation energy over the interval. The time-averaged wavelet spectrogram ($PSD_{WL}$) is shown as the orange line in Fig.~\ref{fig:event-p1}\textbf{(b)}, serving as a smoothed version of $PSD_{FFT}$. The rectified $PSD_{WL}\cdot f$ is shown as the orange dashed line, and is also plotted in linear-log scale on the right y-axis (black line). This approach leverages the geometric interpretation of $PSD_{WL}\cdot f$, as the area under the curve directly reflects the distribution of fluctuation energy over logarithmic frequencies. A significant peak is found at $1/f_{peak} = 299\,\mathrm{s}$ (black dashed line in both Fig.~\ref{fig:event-p1}\textbf{(b)} and Fig.~\ref{fig:event-p2}\textbf{(b)}), corresponding to the 5-minute oscillation signal.

Notably, the 5-minute peak appears at the low-frequency end of the $1/f$ plateau, which extends up to about 20 mHz (50 s). We identify the frequency range containing the central 50\% of the total fluctuation energy, highlighted by the green shaded area. The midpoint of this range is $1/f_{mid} = 105\,\mathrm{s}$ (red dashed line in both Fig.~\ref{fig:event-p1}\textbf{(b)} and Fig.~\ref{fig:event-p2}\textbf{(b)}). During this interval, PSP's spacecraft radial velocity ($\sim 8.4\,\mathrm{km/s}$) was negligible compared to the group velocity of Alfvén waves ($V_r + V_A \sim 915\,\mathrm{km/s}$), and the Alfvénicity remained high ($|\sigma_c| \sim 1$). Therefore, the PSD measured by the spacecraft can be directly mapped to the PSD in the lower corona. Note that $1/f_{mid} = 105\,\mathrm{s}$ is consistent with previous studies that found a concentration of fluctuation energy near a 2-minute period close to the Alfv\'en surface \cite{huang_dominance_2024,huang_two_2025}.

\subsection*{5-Minute Oscillation Properties}

To further characterize the event, we examine its detailed properties in Fig.~\ref{fig:sigma_c}. The frequency-dependent cross helicity $\sigma_c(f)$ and its smoothed version are shown in panel \textbf{(a)}. Panel \textbf{(b)} presents the rectified trace magnetic FFT PSD $P(f)\cdot f$ and the time-averaged wavelet spectrogram $\langle S(f,t)\cdot f\rangle$ in linear-log scale. A prominent peak is observed at $3.2\,\mathrm{mHz}$ ($\approx 1/312\,\mathrm{s}^{-1}$) in $P(f)\cdot f$. To assess its significance, we compute the mean and standard deviation of $P(f)\cdot f$ within the frequency range containing the central 50\% of the total fluctuation energy (green shaded area). Relative to this baseline, the peak at 3.2 mHz rises $5.8\sigma$ above the mean. The frequency-dependent cross helicity $\sigma_c(f)$ further confirms that the fluctuations are highly Alfv\'enic across the frequencies where most of the fluctuation energy resides (see \cite{methods} for parameter definitions).

As shown in Fig.~\ref{fig:event-p2}\textbf{(b)}, the 5-minute oscillation signal consists of three distinct wave packets. To illustrate this, Fig.~\ref{fig:sigma_c} panels \textbf{(c)} to \textbf{(i)} present the rectified magnetic wavelet spectrograms for the three components (R-T-N) and their trace (see \cite{methods} for coordinate definitions). Most of the 5-minute oscillation energy is concentrated in two short-lived wave packets around 10:30 and 11:00, followed by a sustained wave train from 11:20 to 11:55. Notably, the majority of the 5-minute signal appears in the $B_n$ component (panel \textbf{(i)}), which is perpendicular to the ecliptic plane. The 1-minute rolling average in panel \textbf{(h)} clearly shows quasi-monochromatic 5-minute oscillations during 11:20–11:55, closely resembling the line-of-sight velocity measurements observed on the photosphere \cite{musman_vertical_1970}.

A closer examination of the wave train from 11:20 to 11:55 is shown in Fig.~\ref{fig:VB}. Panels \textbf{(b)} and \textbf{(f-h)} display high-resolution magnetic field measurements ($\sim$288~Hz), while panels \textbf{(c-e)} present the magnetic field and proton velocity resampled to a 2-second cadence. All three components of the magnetic field and proton velocity exhibit strong correlation, particularly $B_n$ and $V_n$. Throughout the event, $|B|$ remains nearly constant, and thus the three components fluctuate on a constant sphere. Combined with the close correlation between $\vec{\textbf{V}}$ and $\vec{\textbf{B}}$, these fluctuations are identified as large-amplitude, outward-propagating, spherically polarized Alfv\'en waves (some of which are referred to as ``switchbacks'' in other studies when $B_r$ undergoes a large reversal; see, e.g., \cite{bale_highly_2019,tenerani_evolution_2021,shi_analytic_2024}). Notably, as shown in Fig.~\ref{fig:sigma_c}\textbf{(c-i)}, even in the presence of strong 5-minute oscillations, the majority of the fluctuation energy originates from the pronounced ``switchback'' near 10:30, with its energy concentrated around $10^{-2}$~Hz.

\subsection*{Interpretation of Spacecraft Timeseries}

Interpreting the PSD measured by PSP as a direct remnant of the PSD at the coronal base requires consideration of two key factors: nonlinear interactions and Doppler-shifted signals from structures in the perpendicular direction. As previously discussed, PSP's radial velocity is negligible compared to the group velocity of the Alfv\'en waves ($V_r + V_A$), effectively placing the spacecraft in the same frame of reference as the solar surface for the radial direction. In this static solar frame, the frequency of Alfv\'en waves remains unchanged from their launch frequency, provided nonlinear effects are insignificant. As argued in \cite{huang_dominance_2024}, the short propagation time of Alfv\'en waves ($\lesssim$2 hours) from the lower corona to the spacecraft near the Alfv\'en surface is likely insufficient for nonlinear effects to significantly shift the wave frequencies.

The effect of perpendicular Doppler shift is more subtle. Consider a periodic structure in the perpendicular ($\phi$) direction (along the local spacecraft trajectory) with wavenumber $k$. Such a spatial structure can be observed as a temporal signal via Doppler shift: $f = k\cdot V_{\phi, PSP}$ (where $V_{\phi, PSP}\sim 170\,\mathrm{km/s}$ in this event). However, it is unlikely that the observed 5-minute oscillation arises from perpendicular Doppler shift. First, the fluctuations at the 5-minute period are highly Alfv\'enic (see Fig.~\ref{fig:sigma_c}\textbf{(a)}), indicating the presence of propagating Alfv\'en waves, whereas spatial structures do not necessarily exhibit magnetic-velocity correlation. Second, the majority of the fluctuation energy is concentrated in the $B_n$ component, with negligible contribution from $B_r$ and $B_t$. If the 5-minute oscillation were due to perpendicular Doppler-shifted structures, strong fluctuation power would not be expected to concentrate in only one of the three components.

The 5-minute oscillation event from E24, shown in Fig.~\ref{fig:sup_E24_event-p1}, Fig.~\ref{fig:sup_E24_event-p2}, and Fig.~\ref{fig:sup_E24_polarization}, exhibits properties very similar to those of the E22 event. A significant peak appears at 3.1~mHz, with a significance of 6.2$\sigma$ above the mean. The event is primarily composed of a 35-minute-long Alfv\'en wave train, with fluctuations concentrated in the $B_n$ component, where the signal resembles a quasi-monochromatic wave train. Notably, for both events, the 5-minute oscillation wave trains are consistent with photospheric observations: both last approximately 35 minutes, in agreement with remote sensing data (see Fig.~5 of \cite{musman_vertical_1970}, where typical wave trains last about 30 minutes). Moreover, during both wave trains, PSP traversed about 3.5 degrees in Carrington longitude (see top x-axis in Fig.~\ref{fig:sigma_c} and Fig.~\ref{fig:sup_E24_polarization}). This suggests that if the observed signals originate from p-modes, the p-mode must remain in-phase over a similar or larger spatial scale on the photosphere. Based on remote sensing, 5-minute oscillations typically remain in-phase over about 0.5 arcmin on the solar disk (solar radius $\sim$16 arcmin; in Fig.~5 of \cite{musman_vertical_1970}, the typical signal remains in phase over a spatial distance of 0.5 arcmin), corresponding to about 1.8 degrees in Carrington longitude (0.5 arcmin/16 arcmin/$\pi \times 180^\circ$). Considering a typical super-radial expansion factor ($A \sim 3$) from the photosphere to the coronal base~\cite{verdini_origin_2012}, the spatial extent of the wave train would fall within the coherent range of a typical 5-minute oscillation wave packet on the Sun~\cite{musman_vertical_1970}.

\vspace{2em}
In conclusion, our observations provide the first definitive \textit{in situ} evidence of 5-minute oscillations in the upper solar corona and primordial solar wind. These results demonstrate that 5-minute oscillations can survive passage through the solar corona and reach into, and potentially influence, the solar wind.

\newpage
\begin{figure} 
	\centering
	\includegraphics[width=1.0\textwidth]{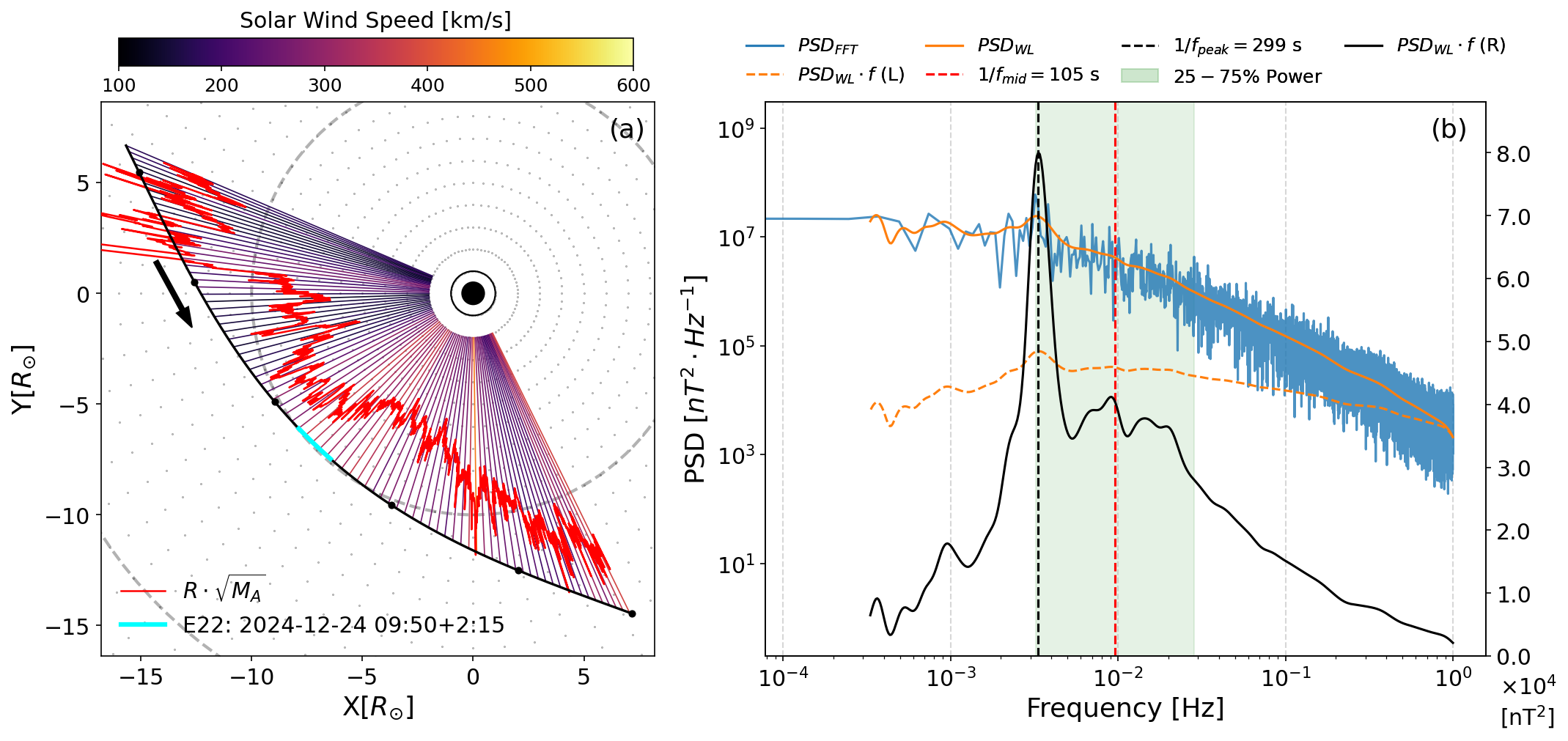} 
	\caption{\textbf{5-Minute Oscillation Event from PSP E22}
		(\textbf{a}) Parker Solar Probe (PSP) trajectory in a co-rotating frame (black). Radial lines show solar wind speed (proton bulk speed); red line shows Alfvén Mach number. Cyan bar marks the selected interval (2024-12-24 09:50–12:05). Black dots are spaced every 8 hours. 
		(\textbf{b}) Trace magnetic power spectral density (PSD): FFT ($PSD_{FFT}$, blue), average wavelet ($PSD_{WL}$, orange), and frequency-rectified PSD ($PSD_{WL} \cdot f$ (L), dashed orange). Right axis: rectified PSD shown in linear-log scale ($PSD_{WL} \cdot f$ (R), black). $f_{peak}$ and $f_{mid}$ denote peak and median frequencies. 
		}
	\label{fig:event-p1} 
\end{figure}

\begin{figure} 
	\centering
	\includegraphics[width=1.0\textwidth]{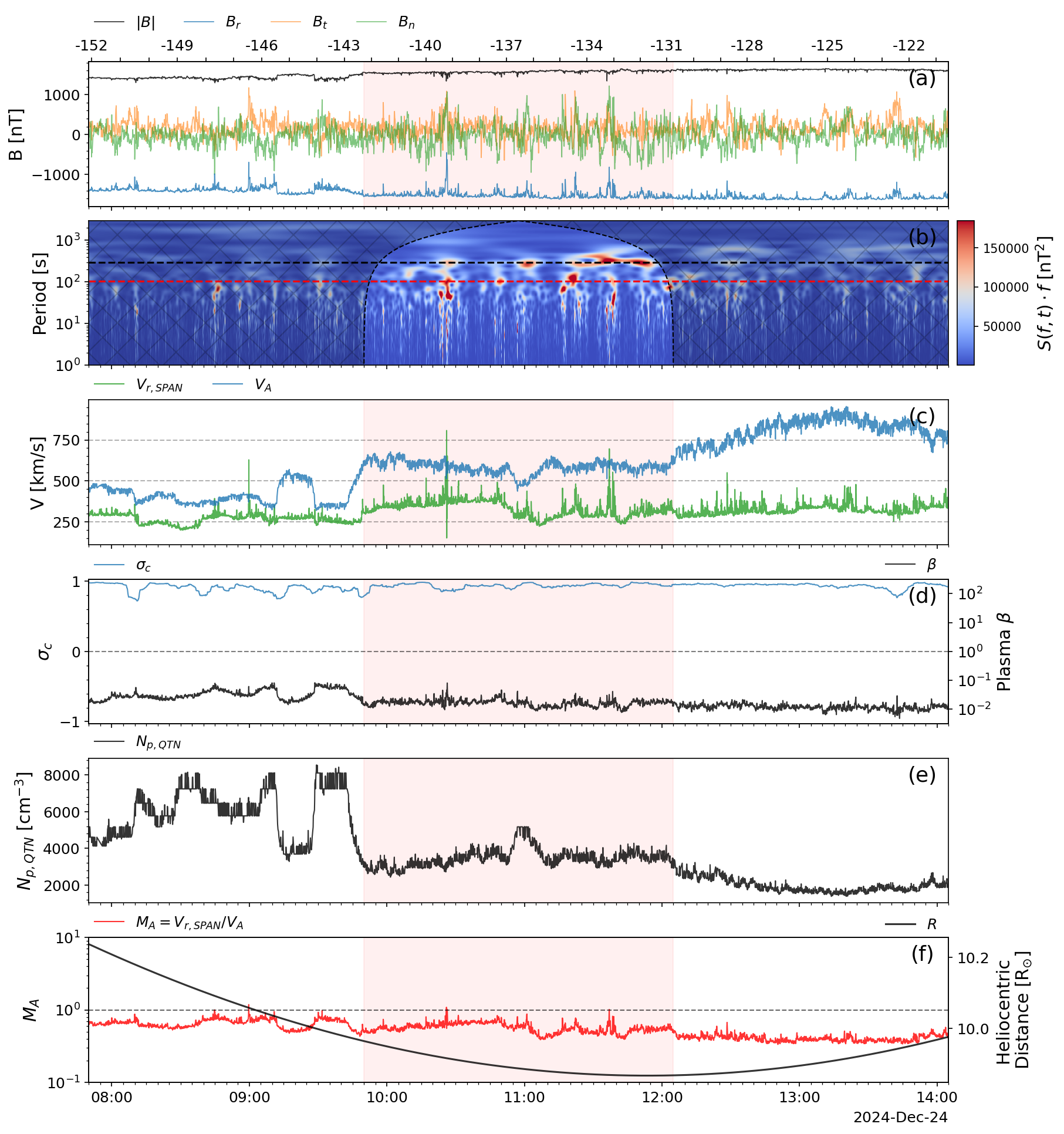} 
	\caption{\textbf{Timeseries of E22 Event}
		(\textbf{a}) Magnetic field time series with red shading indicating the selected interval. Carrington longitude shown above. 
		(\textbf{b}) Frequency-rectified magnetic wavelet spectrogram $S(f,t)\cdot f$, with cone of influence (black dashed curve and shaded region). Horizontal dashed lines mark $1/f_{mid}$ and $1/f_{peak}$. 
		(\textbf{c}) Radial bulk speeds: protons (green) and local Alfvén speed (blue). 
		(\textbf{d}) Cross helicity $\sigma_c$ and plasma $\beta$. 
		(\textbf{e}) Proton density from QTN. 
		(\textbf{f}) PSP’s heliocentric distance and Alfvén Mach number.
		}
	\label{fig:event-p2} 
\end{figure}

\begin{figure} 
	\centering
	\includegraphics[width=1.0\textwidth]{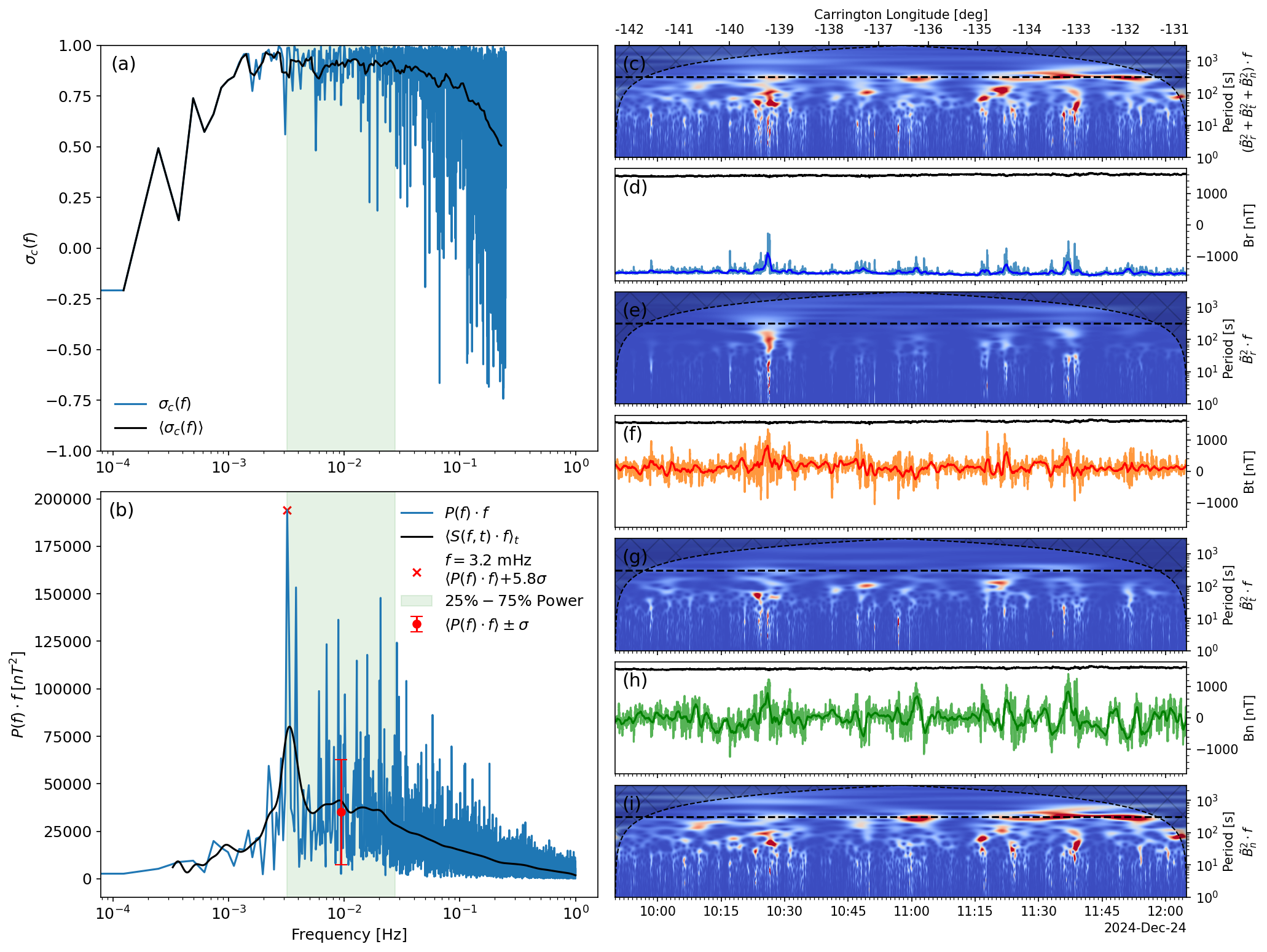} 
	\caption{\textbf{Oscillation Polarization}
	(\textbf{a}) Frequency-dependent cross helicity $\sigma_c(f)$. 
	(\textbf{b}) Magnetic field power spectral density: FFT (blue) and wavelet average (orange). Green shading marks the frequency band containing the central 50\% of fluctuation energy. 
	(\textbf{c}) Magnetic wavelet spectrogram with 5-minute period indicated (green dashed line). 
	(\textbf{d}) Magnetic field strength $|B|$, radial component $B_r$, and 1-minute average of $B_r$. 
	(\textbf{e}) Wavelet spectrogram of $B_r$. 
	(\textbf{f}) $|B|$, tangential component $B_t$, and 1-minute average of $B_t$. 
	(\textbf{g}) Wavelet spectrogram of $B_t$. 
	(\textbf{h}) $|B|$, normal component $B_n$, and 1-minute average of $B_n$. 
	(\textbf{i}) Wavelet spectrogram of $B_n$.
	}
	\label{fig:sigma_c} 
\end{figure}

\begin{figure} 
	\centering
	\includegraphics[width=1.0\textwidth]{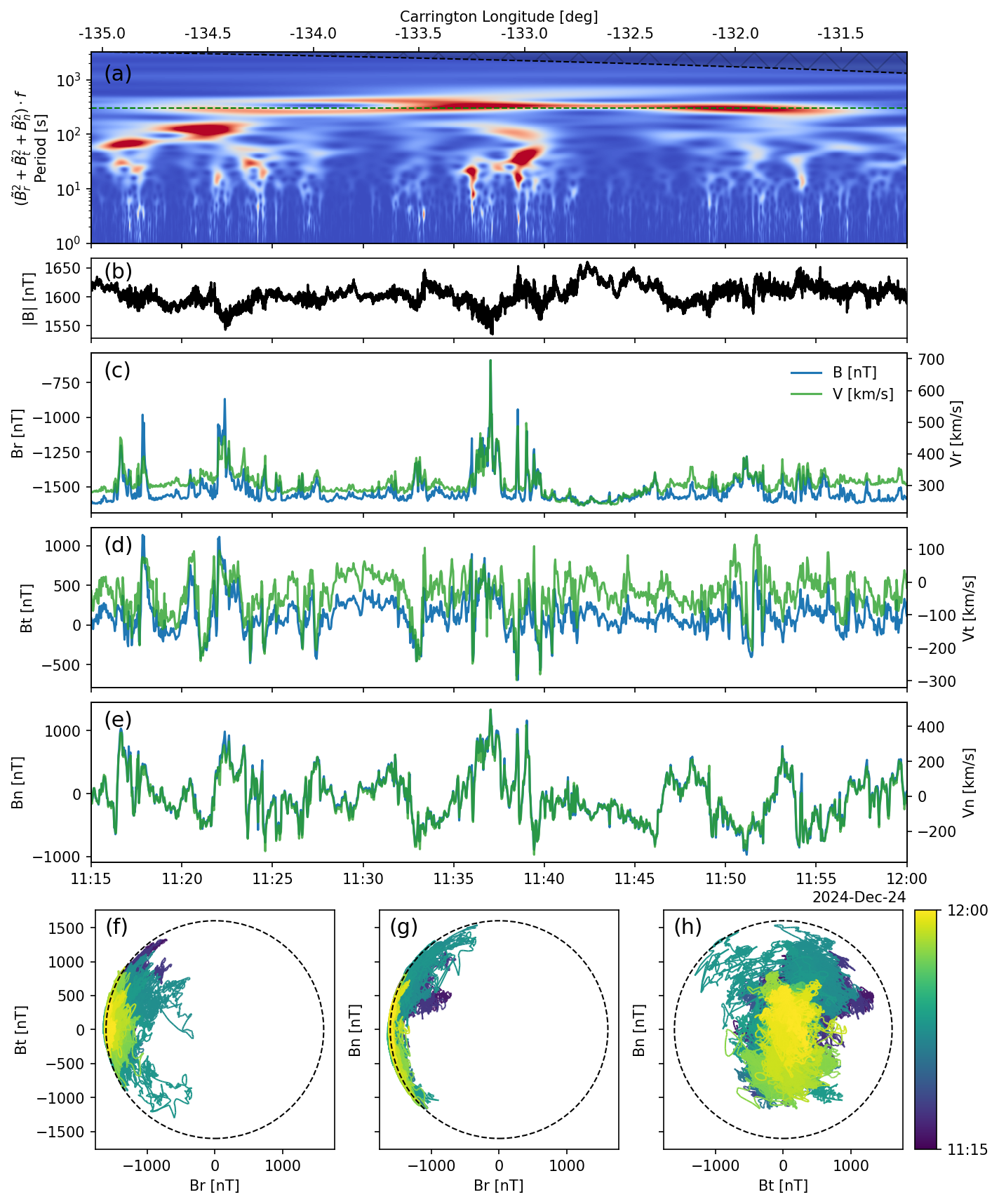} 
	\caption{\textbf{Spherical Polarization of the Oscillations}
	(\textbf{a}) Trace magnetic periodogram. 5-minute is highlighted with the green dashed line. (\textbf{b}) $|B|$. (\textbf{c}) $B_r$ and $V_r$. (\textbf{d}) $B_t$ and $V_r$. (\textbf{e}) $B_n$ and $V_n$. (\textbf{f}) Hodogram of $B_t-B_r$. (\textbf{g}) $B_n-B_r$. (\textbf{h}) $B_n-B_t$.}
	\label{fig:VB} 
\end{figure}


\clearpage 

%
\bibliography{science_template} 

@Misc{methods,
  note = {Materials and methods are available as supplementary material},
}

@article{cally_benchmarking_2011,
	title = {{BENCHMARKING} {FAST}-{TO}-{ALFV{\'E}N} {MODE} {CONVERSION} {IN} {A} {COLD} {MAGNETOHYDRODYNAMIC} {PLASMA}},
	volume = {738},
	issn = {0004-637X},
	url = {https://dx.doi.org/10.1088/0004-637X/738/2/119},
	doi = {10.1088/0004-637X/738/2/119},
	language = {en},
	number = {2},
	urldate = {2024-04-29},
	journal = {The Astrophysical Journal},
	author = {Cally, Paul S. and Hansen, Shelley C.},
	month = aug,
	year = {2011},
	note = {Publisher: The American Astronomical Society},
	pages = {119},
}

@article{hansen_benchmarking_2012,
	title = {{BENCHMARKING} {FAST}-{TO}-{ALFV{\'E}N} {MODE} {CONVERSION} {IN} {A} {COLD} {MHD} {PLASMA}. {II}. {HOW} {TO} {GET} {ALFV{\'E}N} {WAVES} {THROUGH} {THE} {SOLAR} {TRANSITION} {REGION}},
	volume = {751},
	issn = {0004-637X},
	url = {https://dx.doi.org/10.1088/0004-637X/751/1/31},
	doi = {10.1088/0004-637X/751/1/31},
	language = {en},
	number = {1},
	urldate = {2024-04-28},
	journal = {The Astrophysical Journal},
	author = {Hansen, Shelley C. and Cally, Paul S.},
	month = may,
	year = {2012},
	note = {Publisher: The American Astronomical Society},
	pages = {31},
}

@article{morton_basal_2019,
	title = {A basal contribution from p-modes to the {Alfv{\'e}nic} wave flux in the {Sun}{\textquoteright}s corona},
	volume = {3},
	issn = {2397-3366},
	url = {https://www.nature.com/articles/s41550-018-0668-9},
	doi = {10.1038/s41550-018-0668-9},
	language = {en},
	number = {3},
	urldate = {2024-01-09},
	journal = {Nature Astronomy},
	author = {Morton, R. J. and Weberg, M. J. and McLaughlin, J. A.},
	month = jan,
	year = {2019},
	pages = {223--229},
}

@article{bruno_solar_2013,
	title = {The {Solar} {Wind} as a {Turbulence} {Laboratory}},
	volume = {10},
	issn = {1614-4961},
	url = {https://doi.org/10.12942/lrsp-2013-2},
	doi = {10.12942/lrsp-2013-2},
	number = {1},
	journal = {Living Reviews in Solar Physics},
	author = {Bruno, Roberto and Carbone, Vincenzo},
	year = {2013},
	pages = {2},
}

@article{fox_solar_2016,
	title = {The {Solar} {Probe} {Plus} {Mission}: {Humanity}{\textquoteright}s {First} {Visit} to {Our} {Star}},
	volume = {204},
	issn = {0038-6308, 1572-9672},
	shorttitle = {The {Solar} {Probe} {Plus} {Mission}},
	url = {http://link.springer.com/10.1007/s11214-015-0211-6},
	doi = {10.1007/s11214-015-0211-6},
	language = {en},
	number = {1-4},
	urldate = {2020-08-09},
	journal = {Space Science Reviews},
	author = {Fox, N. J. and Velli, M. C. and Bale, S. D. and Decker, R. and Driesman, A. and Howard, R. A. and Kasper, J. C. and Kinnison, J. and Kusterer, M. and Lario, D. and Lockwood, M. K. and McComas, D. J. and Raouafi, N. E. and Szabo, A.},
	month = dec,
	year = {2016},
	pages = {7--48},
}

@article{kasper_solar_2016,
	title = {Solar {Wind} {Electrons} {Alphas} and {Protons} ({SWEAP}) {Investigation}: {Design} of the {Solar} {Wind} and {Coronal} {Plasma} {Instrument} {Suite} for {Solar} {Probe} {Plus}},
	volume = {204},
	doi = {10.1007/s11214-015-0206-3},
	number = {1-4},
	journal = {{\ss}r},
	author = {Kasper, Justin C. and Abiad, Robert and Austin, Gerry and Balat-Pichelin, Marianne and Bale, Stuart D. and Belcher, John W. and Berg, Peter and Bergner, Henry and Berthomier, Matthieu and Bookbinder, Jay and Brodu, Etienne and Caldwell, David and Case, Anthony W. and Chandran, Benjamin D. G. and Cheimets, Peter and Cirtain, Jonathan W. and Cranmer, Steven R. and Curtis, David W. and Daigneau, Peter and Dalton, Greg and Dasgupta, Brahmananda and DeTomaso, David and Diaz-Aguado, Millan and Djordjevic, Blagoje and Donaskowski, Bill and Effinger, Michael and Florinski, Vladimir and Fox, Nichola and Freeman, Mark and Gallagher, Dennis and Gary, S. Peter and Gauron, Tom and Gates, Richard and Goldstein, Melvin and Golub, Leon and Gordon, Dorothy A. and Gurnee, Reid and Guth, Giora and Halekas, Jasper and Hatch, Ken and Heerikuisen, Jacob and Ho, George and Hu, Qiang and Johnson, Greg and Jordan, Steven P. and Korreck, Kelly E. and Larson, Davin and Lazarus, Alan J. and Li, Gang and Livi, Roberto and Ludlam, Michael and Maksimovic, Milan and McFadden, James P. and Marchant, William and Maruca, Bennet A. and McComas, David J. and Messina, Luciana and Mercer, Tony and Park, Sang and Peddie, Andrew M. and Pogorelov, Nikolai and Reinhart, Matthew J. and Richardson, John D. and Robinson, Miles and Rosen, Irene and Skoug, Ruth M. and Slagle, Amanda and Steinberg, John T. and Stevens, Michael L. and Szabo, Adam and Taylor, Ellen R. and Tiu, Chris and Turin, Paul and Velli, Marco and Webb, Gary and Whittlesey, Phyllis and Wright, Ken and Wu, S. T. and Zank, Gary},
	month = dec,
	year = {2016},
	pages = {131--186},
}

@article{bale_fields_2016,
	title = {The {FIELDS} {Instrument} {Suite} for {Solar} {Probe} {Plus}. {Measuring} the {Coronal} {Plasma} and {Magnetic} {Field}, {Plasma} {Waves} and {Turbulence}, and {Radio} {Signatures} of {Solar} {Transients}},
	volume = {204},
	doi = {10.1007/s11214-016-0244-5},
	number = {1-4},
	journal = {{\ss}r},
	author = {Bale, S. D. and Goetz, K. and Harvey, P. R. and Turin, P. and Bonnell, J. W. and Dudok de Wit, T. and Ergun, R. E. and MacDowall, R. J. and Pulupa, M. and Andre, M. and Bolton, M. and Bougeret, J. -L. and Bowen, T. A. and Burgess, D. and Cattell, C. A. and Chandran, B. D. G. and Chaston, C. C. and Chen, C. H. K. and Choi, M. K. and Connerney, J. E. and Cranmer, S. and Diaz-Aguado, M. and Donakowski, W. and Drake, J. F. and Farrell, W. M. and Fergeau, P. and Fermin, J. and Fischer, J. and Fox, N. and Glaser, D. and Goldstein, M. and Gordon, D. and Hanson, E. and Harris, S. E. and Hayes, L. M. and Hinze, J. J. and Hollweg, J. V. and Horbury, T. S. and Howard, R. A. and Hoxie, V. and Jannet, G. and Karlsson, M. and Kasper, J. C. and Kellogg, P. J. and Kien, M. and Klimchuk, J. A. and Krasnoselskikh, V. V. and Krucker, S. and Lynch, J. J. and Maksimovic, M. and Malaspina, D. M. and Marker, S. and Martin, P. and Martinez-Oliveros, J. and McCauley, J. and McComas, D. J. and McDonald, T. and Meyer-Vernet, N. and Moncuquet, M. and Monson, S. J. and Mozer, F. S. and Murphy, S. D. and Odom, J. and Oliverson, R. and Olson, J. and Parker, E. N. and Pankow, D. and Phan, T. and Quataert, E. and Quinn, T. and Ruplin, S. W. and Salem, C. and Seitz, D. and Sheppard, D. A. and Siy, A. and Stevens, K. and Summers, D. and Szabo, A. and Timofeeva, M. and Vaivads, A. and Velli, M. and Yehle, A. and Werthimer, D. and Wygant, J. R.},
	month = dec,
	year = {2016},
	pages = {49--82},
}

@article{shi_acceleration_2022,
	title = {Acceleration of polytropic solar wind: {Parker} {Solar} {Probe} observation and one-dimensional model},
	volume = {29},
	issn = {1070-664X},
	shorttitle = {Acceleration of polytropic solar wind},
	url = {https://aip.scitation.org/doi/10.1063/5.0124703},
	doi = {10.1063/5.0124703},
	number = {12},
	urldate = {2023-02-12},
	journal = {Physics of Plasmas},
	author = {Shi, Chen (??) and Velli, Marco and Bale, Stuart D. and R{\'e}ville, Victor and Maksimovi{\'c}, Milan and Dakeyo, Jean-Baptiste},
	month = dec,
	year = {2022},
	note = {Publisher: American Institute of Physics},
	pages = {122901},
}

@article{cally_alfven_2012,
	title = {Alfv{\'e}n {Reflection} and {Reverberation} in the {Solar} {Atmosphere}},
	volume = {280},
	issn = {1573-093X},
	url = {https://doi.org/10.1007/s11207-012-0052-3},
	doi = {10.1007/s11207-012-0052-3},
	language = {en},
	number = {1},
	urldate = {2023-06-06},
	journal = {Solar Physics},
	author = {Cally, P. S.},
	month = sep,
	year = {2012},
	pages = {33--50},
}

@article{belcher_alfvenic_1971,
	title = {{ALFV{\'E}NIC} {Wave} {Pressures} and the {Solar} {Wind}},
	doi = {10.1086/151105},
	journal = {Astrophysical Journal, vol. 168, p.509},
	author = {Belcher, J. W.},
	year = {1971},
}

@article{panasenco_exploring_2020,
	title = {Exploring {Solar} {Wind} {Origins} and {Connecting} {Plasma} {Flows} from the {Parker} {Solar} {Probe} to 1 au: {Nonspherical} {Source} {Surface} and {Alfv{\'e}nic} {Fluctuations}},
	volume = {246},
	issn = {0067-0049},
	shorttitle = {Exploring {Solar} {Wind} {Origins} and {Connecting} {Plasma} {Flows} from the {Parker} {Solar} {Probe} to 1 au},
	url = {https://dx.doi.org/10.3847/1538-4365/ab61f4},
	doi = {10.3847/1538-4365/ab61f4},
	language = {en},
	number = {2},
	urldate = {2023-03-21},
	journal = {The Astrophysical Journal Supplement Series},
	author = {Panasenco, Olga and Velli, Marco and D{\textquoteright}Amicis, Raffaella and Shi, Chen and R{\'e}ville, Victor and Bale, Stuart D. and Badman, Samuel T. and Kasper, Justin and Korreck, Kelly and Bonnell, J. W. and Wit, Thierry Dudok de and Goetz, Keith and Harvey, Peter R. and MacDowall, Robert J. and Malaspina, David M. and Pulupa, Marc and Case, Anthony W. and Larson, Davin and Livi, Roberto and Stevens, Michael and Whittlesey, Phyllis},
	month = feb,
	year = {2020},
	note = {Publisher: The American Astronomical Society},
	pages = {54},
}

@article{kasper_parker_2021,
	title = {Parker {Solar} {Probe} {Enters} the {Magnetically} {Dominated} {Solar} {Corona}},
	volume = {127},
	url = {https://link.aps.org/doi/10.1103/PhysRevLett.127.255101},
	doi = {10.1103/PhysRevLett.127.255101},
	number = {25},
	urldate = {2022-09-05},
	journal = {Physical Review Letters},
	author = {Kasper, J. C. and Klein, K. G. and Lichko, E. and Huang, Jia and Chen, C. H. K. and Badman, S. T. and Bonnell, J. and Whittlesey, P. L. and Livi, R. and Larson, D. and Pulupa, M. and Rahmati, A. and Stansby, D. and Korreck, K. E. and Stevens, M. and Case, A. W. and Bale, S. D. and Maksimovic, M. and Moncuquet, M. and Goetz, K. and Halekas, J. S. and Malaspina, D. and Raouafi, Nour E. and Szabo, A. and MacDowall, R. and Velli, Marco and Dudok de Wit, Thierry and Zank, G. P.},
	month = dec,
	year = {2021},
	note = {Publisher: American Physical Society},
	pages = {255101},
}

@article{velli_propagation_1993,
	title = {On the propagation of ideal, linear {Alfven} waves in radially stratified stellar atmospheres and winds},
	volume = {270},
	issn = {0004-6361},
	url = {https://ui.adsabs.harvard.edu/abs/1993A&A...270..304V},
	urldate = {2023-04-13},
	journal = {Astronomy and Astrophysics},
	author = {Velli, M.},
	month = mar,
	year = {1993},
	note = {ADS Bibcode: 1993A\&A...270..304V},
	pages = {304--314},
}

@article{heinemann_non-wkb_1980,
	title = {Non-{WKB} {Alfv{\'e}n} waves in the solar wind},
	volume = {85},
	issn = {01480227},
	url = {http://doi.wiley.com/10.1029/JA085iA03p01311},
	doi = {10.1029/JA085iA03p01311},
	language = {en},
	number = {A3},
	urldate = {2021-09-20},
	journal = {Journal of Geophysical Research: Space Physics},
	author = {Heinemann, M. and Olbert, S.},
	month = mar,
	year = {1980},
	pages = {1311--1327},
}

@article{verdini_origin_2012,
	title = {{ON} {THE} {ORIGIN} {OF} {THE} 1/ \textit{f} {SPECTRUM} {IN} {THE} {SOLAR} {WIND} {MAGNETIC} {FIELD}},
	volume = {750},
	issn = {2041-8205, 2041-8213},
	url = {https://iopscience.iop.org/article/10.1088/2041-8205/750/2/L33},
	doi = {10.1088/2041-8205/750/2/L33},
	language = {en},
	number = {2},
	urldate = {2023-02-19},
	journal = {The Astrophysical Journal},
	author = {Verdini, Andrea and Grappin, Roland and Pinto, Rui and Velli, Marco},
	month = may,
	year = {2012},
	pages = {L33},
}

@article{harris_array_2020,
	title = {Array programming with {NumPy}},
	volume = {585},
	copyright = {2020 The Author(s)},
	issn = {1476-4687},
	url = {https://www.nature.com/articles/s41586-020-2649-2},
	doi = {10.1038/s41586-020-2649-2},
	language = {en},
	number = {7825},
	urldate = {2023-02-17},
	journal = {Nature},
	author = {Harris, Charles R. and Millman, K. Jarrod and van der Walt, St{\'e}fan J. and Gommers, Ralf and Virtanen, Pauli and Cournapeau, David and Wieser, Eric and Taylor, Julian and Berg, Sebastian and Smith, Nathaniel J. and Kern, Robert and Picus, Matti and Hoyer, Stephan and van Kerkwijk, Marten H. and Brett, Matthew and Haldane, Allan and del R{\'i}o, Jaime Fern{\'a}ndez and Wiebe, Mark and Peterson, Pearu and G{\'e}rard-Marchant, Pierre and Sheppard, Kevin and Reddy, Tyler and Weckesser, Warren and Abbasi, Hameer and Gohlke, Christoph and Oliphant, Travis E.},
	month = sep,
	year = {2020},
	note = {Number: 7825
Publisher: Nature Publishing Group},
	pages = {357--362},
}

@article{livi_solar_2022,
	title = {The {Solar} {Probe} {ANalyzer}{\textemdash}{Ions} on the {Parker} {Solar} {Probe}},
	volume = {938},
	issn = {0004-637X, 1538-4357},
	url = {https://iopscience.iop.org/article/10.3847/1538-4357/ac93f5},
	doi = {10.3847/1538-4357/ac93f5},
	number = {2},
	urldate = {2023-02-14},
	journal = {The Astrophysical Journal},
	author = {Livi, Roberto and Larson, Davin E. and Kasper, Justin C. and Abiad, Robert and Case, A. W. and Klein, Kristopher G. and Curtis, David W. and Dalton, Gregory and Stevens, Michael and Korreck, Kelly E. and Ho, George and Robinson, Miles and Tiu, Chris and Whittlesey, Phyllis L. and Verniero, Jaye L. and Halekas, Jasper and McFadden, James and Marckwordt, Mario and Slagle, Amanda and Abatcha, Mamuda and Rahmati, Ali and McManus, Michael D.},
	month = oct,
	year = {2022},
	pages = {138},
}

@article{bale_highly_2019,
	title = {Highly structured slow solar wind emerging from an equatorial coronal hole},
	copyright = {2019 The Author(s), under exclusive licence to Springer Nature Limited},
	issn = {1476-4687},
	url = {https://www.nature.com/articles/s41586-019-1818-7},
	doi = {10.1038/s41586-019-1818-7},
	language = {en},
	urldate = {2019-12-06},
	journal = {Nature},
	author = {Bale, S. D. and Badman, S. T. and Bonnell, J. W. and Bowen, T. A. and Burgess, D. and Case, A. W. and Cattell, C. A. and Chandran, B. D. G. and Chaston, C. C. and Chen, C. H. K. and Drake, J. F. and de Wit, T. Dudok and Eastwood, J. P. and Ergun, R. E. and Farrell, W. M. and Fong, C. and Goetz, K. and Goldstein, M. and Goodrich, K. A. and Harvey, P. R. and Horbury, T. S. and Howes, G. G. and Kasper, J. C. and Kellogg, P. J. and Klimchuk, J. A. and Korreck, K. E. and Krasnoselskikh, V. V. and Krucker, S. and Laker, R. and Larson, D. E. and MacDowall, R. J. and Maksimovic, M. and Malaspina, D. M. and Martinez-Oliveros, J. and McComas, D. J. and Meyer-Vernet, N. and Moncuquet, M. and Mozer, F. S. and Phan, T. D. and Pulupa, M. and Raouafi, N. E. and Salem, C. and Stansby, D. and Stevens, M. and Szabo, A. and Velli, M. and Woolley, T. and Wygant, J. R.},
	month = dec,
	year = {2019},
	pages = {1--6},
}

@article{shi_analytic_2024,
	title = {Analytic {Model} and {Magnetohydrodynamic} {Simulations} of {Three}-dimensional {Magnetic} {Switchbacks}},
	volume = {964},
	issn = {2041-8205},
	url = {https://dx.doi.org/10.3847/2041-8213/ad335a},
	doi = {10.3847/2041-8213/ad335a},
	language = {en},
	number = {2},
	urldate = {2024-04-19},
	journal = {The Astrophysical Journal Letters},
	author = {Shi, Chen and Velli, Marco and Toth, Gabor and Zhang, Kun and Tenerani, Anna and Huang, Zesen and Sioulas, Nikos and Holst, Bart van der},
	month = mar,
	year = {2024},
	note = {Publisher: The American Astronomical Society},
	pages = {L28},
}

@article{huang_conservation_2022,
	title = {Conservation of {Total} {Wave} {Action} in the {Expanding} {Solar} {Wind}},
	volume = {935},
	copyright = {All rights reserved},
	issn = {0004-637X},
	url = {https://doi.org/10.3847/1538-4357/ac74c5},
	doi = {10.3847/1538-4357/ac74c5},
	language = {en},
	number = {1},
	urldate = {2022-09-04},
	journal = {The Astrophysical Journal},
	author = {Huang, Zesen and Shi, Chen and Sioulas, Nikos and Velli, Marco},
	month = aug,
	year = {2022},
	note = {Publisher: American Astronomical Society},
	pages = {60},
}

@article{davis_evolution_2023,
	title = {The {Evolution} of the 1/f {Range} within a {Single} {Fast}-solar-wind {Stream} between 17.4 and 45.7 {Solar} {Radii}},
	volume = {950},
	issn = {0004-637X},
	url = {https://dx.doi.org/10.3847/1538-4357/acd177},
	doi = {10.3847/1538-4357/acd177},
	language = {en},
	number = {2},
	urldate = {2024-02-04},
	journal = {The Astrophysical Journal},
	author = {Davis, Nooshin and Chandran, B. D. G. and Bowen, T. A. and Badman, S. T. and Wit, T. Dudok de and Chen, C. H. K. and Bale, S. D. and Huang, Zesen and Sioulas, Nikos and Velli, Marco},
	month = jun,
	year = {2023},
	note = {Publisher: The American Astronomical Society},
	pages = {154},
}

@article{huang_new_2023,
	title = {New {Observations} of {Solar} {Wind} 1/f {Turbulence} {Spectrum} from {Parker} {Solar} {Probe}},
	volume = {950},
	issn = {2041-8205},
	url = {https://dx.doi.org/10.3847/2041-8213/acd7f2},
	doi = {10.3847/2041-8213/acd7f2},
	language = {en},
	number = {1},
	urldate = {2023-06-13},
	journal = {The Astrophysical Journal Letters},
	author = {Huang, Zesen and Sioulas, Nikos and Shi, Chen and Velli, Marco and Bowen, Trevor and Davis, Nooshin and Chandran, B. D. G. and Matteini, Lorenzo and Kang, Ning and Shi, Xiaofei and Huang, Jia and Bale, Stuart D. and Kasper, J. C. and Larson, Davin E. and Livi, Roberto and Whittlesey, P. L. and Rahmati, Ali and Paulson, Kristoff and Stevens, M. and Case, A. W. and Wit, Thierry Dudok de and Malaspina, David M. and Bonnell, J. W. and Goetz, Keith and Harvey, Peter R. and MacDowall, Robert J.},
	month = jun,
	year = {2023},
	note = {Publisher: The American Astronomical Society},
	pages = {L8},
}

@article{leibacher_new_1971,
	title = {{NEW} {DESCRIPTION} {OF} {THE} {SOLAR} {FIVE}-{MINUTE} {OSCILLATION}.},
	url = {https://www.osti.gov/biblio/4067356},
	language = {English},
	urldate = {2024-05-09},
	journal = {Astrophys. Lett. 7: 191-2(Jan 1971).},
	author = {Leibacher, J. and Stein, R. F.},
	month = jan,
	year = {1971},
	note = {Institution: Harvard Coll. Observatory, Cambridge, Mass.},
}

@article{simon_velocity_1964,
	title = {Velocity {Fields} in the {Solar} {Atmosphere}. {III}. {Large}-{Scale} {Motions}, the {Chromospheric} {Network}, and {Magnetic} {Fields}.},
	volume = {140},
	issn = {0004-637X},
	url = {https://ui.adsabs.harvard.edu/abs/1964ApJ...140.1120S},
	doi = {10.1086/148010},
	urldate = {2024-05-04},
	journal = {The Astrophysical Journal},
	author = {Simon, G. W. and Leighton, R. B.},
	month = oct,
	year = {1964},
	note = {Publisher: IOP
ADS Bibcode: 1964ApJ...140.1120S},
	pages = {1120},
}

@article{noyes_velocity_1963,
	title = {Velocity {Fields} in the {Solar} {Atmosphere}. {II}. {The} {Oscillatory} {Field}.},
	volume = {138},
	issn = {0004-637X},
	url = {https://ui.adsabs.harvard.edu/abs/1963ApJ...138..631N},
	doi = {10.1086/147675},
	urldate = {2024-05-04},
	journal = {The Astrophysical Journal},
	author = {Noyes, Robert W. and Leighton, Robert B.},
	month = oct,
	year = {1963},
	note = {Publisher: IOP
ADS Bibcode: 1963ApJ...138..631N},
	pages = {631},
}

@article{leighton_velocity_1962,
	title = {Velocity {Fields} in the {Solar} {Atmosphere}. {I}. {Preliminary} {Report}.},
	volume = {135},
	issn = {0004-637X},
	url = {https://ui.adsabs.harvard.edu/abs/1962ApJ...135..474L},
	doi = {10.1086/147285},
	urldate = {2024-05-04},
	journal = {The Astrophysical Journal},
	author = {Leighton, Robert B. and Noyes, Robert W. and Simon, George W.},
	month = mar,
	year = {1962},
	note = {Publisher: IOP
ADS Bibcode: 1962ApJ...135..474L},
	pages = {474},
}

@article{ulrich_five-minute_1970,
	title = {The {Five}-{Minute} {Oscillations} on the {Solar} {Surface}},
	volume = {162},
	issn = {0004-637X},
	url = {https://ui.adsabs.harvard.edu/abs/1970ApJ...162..993U},
	doi = {10.1086/150731},
	urldate = {2024-04-30},
	journal = {The Astrophysical Journal},
	author = {Ulrich, Roger K.},
	month = dec,
	year = {1970},
	note = {Publisher: IOP
ADS Bibcode: 1970ApJ...162..993U},
	pages = {993},
}

@article{telloni_energy_2023,
	title = {Energy {Budget} in the {Solar} {Corona}},
	volume = {954},
	issn = {0004-637X},
	url = {https://dx.doi.org/10.3847/1538-4357/aceb64},
	doi = {10.3847/1538-4357/aceb64},
	language = {en},
	number = {2},
	urldate = {2023-10-23},
	journal = {The Astrophysical Journal},
	author = {Telloni, Daniele and Romoli, Marco and Velli, Marco and Zank, Gary P. and Adhikari, Laxman and Zhao, Lingling and Downs, Cooper and Halekas, Jasper S. and Verniero, Jaye L. and McManus, Michael D. and Shi, Chen and Burtovoi, Aleksandr and Susino, Roberto and Spadaro, Daniele and Liberatore, Alessandro and Antonucci, Ester and Leo, Yara De and Abbo, Lucia and Frassati, Federica and Jerse, Giovanna and Landini, Federico and Nicolini, Gianalfredo and Pancrazzi, Maurizio and Russano, Giuliana and Sasso, Clementina and Andretta, Vincenzo and Deppo, Vania Da and Fineschi, Silvano and Grimani, Catia and Heinzel, Petr and Moses, John D. and Naletto, Giampiero and Stangalini, Marco and Teriaca, Luca and Uslenghi, Michela and Bale, Stuart D. and Kasper, Justin C.},
	month = aug,
	year = {2023},
	note = {Publisher: The American Astronomical Society},
	pages = {108},
}

@article{bowen_merged_2020,
	title = {A {Merged} {Search}-{Coil} and {Fluxgate} {Magnetometer} {Data} {Product} for {Parker} {Solar} {Probe} {FIELDS}},
	volume = {125},
	copyright = {{\textcopyright}2020. American Geophysical Union. All Rights Reserved.},
	issn = {2169-9402},
	url = {https://onlinelibrary.wiley.com/doi/abs/10.1029/2020JA027813},
	doi = {10.1029/2020JA027813},
	language = {en},
	number = {5},
	urldate = {2023-08-09},
	journal = {Journal of Geophysical Research: Space Physics},
	author = {Bowen, T. A. and Bale, S. D. and Bonnell, J. W. and Dudok de Wit, T. and Goetz, K. and Goodrich, K. and Gruesbeck, J. and Harvey, P. R. and Jannet, G. and Koval, A. and MacDowall, R. J. and Malaspina, D. M. and Pulupa, M. and Revillet, C. and Sheppard, D. and Szabo, A.},
	year = {2020},
	note = {\_eprint: https://onlinelibrary.wiley.com/doi/pdf/10.1029/2020JA027813},
	pages = {e2020JA027813},
}

@article{tenerani_evolution_2021,
	title = {Evolution of {Switchbacks} in the {Inner} {Heliosphere}},
	volume = {919},
	issn = {2041-8205},
	url = {https://doi.org/10.3847/2041-8213/ac2606},
	doi = {10.3847/2041-8213/ac2606},
	language = {en},
	number = {2},
	urldate = {2021-11-22},
	journal = {The Astrophysical Journal Letters},
	author = {Tenerani, Anna and Sioulas, Nikos and Matteini, Lorenzo and Panasenco, Olga and Shi, Chen and Velli, Marco},
	month = oct,
	year = {2021},
	note = {Publisher: American Astronomical Society},
	pages = {L31},
}

@article{morton_alfvenic_2023,
	title = {Alfv{\'e}nic waves in the inhomogeneous solar atmosphere},
	volume = {7},
	issn = {2367-3192},
	url = {https://doi.org/10.1007/s41614-023-00118-3},
	doi = {10.1007/s41614-023-00118-3},
	language = {en},
	number = {1},
	urldate = {2024-01-05},
	journal = {Reviews of Modern Plasma Physics},
	author = {Morton, R. J. and Sharma, R. and Tajfirouze, E. and Miriyala, H.},
	month = mar,
	year = {2023},
	pages = {17},
}

@article{zhao_analysis_2021,
	title = {Analysis of {Magnetohydrodynamic} {Perturbations} in the {Radial}-field {Solar} {Wind} from {Parker} {Solar} {Probe} {Observations}},
	volume = {923},
	url = {https://iopscience.iop.org/article/10.3847/1538-4357/ac2ffe/meta},
	number = {2},
	urldate = {2023-12-16},
	journal = {The Astrophysical Journal},
	author = {Zhao, S. Q. and Yan, Huirong and Liu, Terry Z. and Liu, Mingzhe and Shi, Mijie},
	year = {2021},
	note = {Publisher: IOP Publishing},
	pages = {253},
}

@article{hollweg_alfven_1973,
	title = {Alfv{\'e}n waves in the solar wind: {Wave} pressure, poynting flux, and angular momentum},
	volume = {78},
	issn = {2156-2202},
	shorttitle = {Alfv{\'e}n waves in the solar wind},
	url = {https://onlinelibrary.wiley.com/doi/abs/10.1029/JA078i019p03643},
	doi = {10.1029/JA078i019p03643},
	language = {en},
	number = {19},
	urldate = {2022-09-05},
	journal = {Journal of Geophysical Research (1896-1977)},
	author = {Hollweg, Joseph V.},
	year = {1973},
	note = {\_eprint: https://onlinelibrary.wiley.com/doi/pdf/10.1029/JA078i019p03643},
	pages = {3643--3652},
}

@article{zheng_solar_2024,
	title = {Solar {Wind} {Density} and {Core} {Temperature} {Derived} from the {PSP} {Quasi}-thermal {Noise} {Measurements}},
	volume = {963},
	issn = {0004-637X},
	url = {https://dx.doi.org/10.3847/1538-4357/ad236d},
	doi = {10.3847/1538-4357/ad236d},
	language = {en},
	number = {2},
	urldate = {2024-09-09},
	journal = {The Astrophysical Journal},
	author = {Zheng, Xianming and Liu, Kaijun and Martinovi{\'c}, Mihailo M. and Pierrard, Viviane and Liu, Mingzhe and He, Qingbao and Cheng, Kun and Liu, Yuqi and Wang, Yan},
	month = mar,
	year = {2024},
	note = {Publisher: The American Astronomical Society},
	pages = {154},
}

@article{mostafavi_alphaproton_2022,
	title = {Alpha{\textendash}{Proton} {Differential} {Flow} of the {Young} {Solar} {Wind}: {Parker} {Solar} {Probe} {Observations}},
	volume = {926},
	issn = {2041-8205},
	shorttitle = {Alpha{\textendash}{Proton} {Differential} {Flow} of the {Young} {Solar} {Wind}},
	url = {https://dx.doi.org/10.3847/2041-8213/ac51e1},
	doi = {10.3847/2041-8213/ac51e1},
	language = {en},
	number = {2},
	urldate = {2024-09-07},
	journal = {The Astrophysical Journal Letters},
	author = {Mostafavi, P. and Allen, R. C. and McManus, M. D. and Ho, G. C. and Raouafi, N. E. and Larson, D. E. and Kasper, J. C. and Bale, S. D.},
	month = feb,
	year = {2022},
	note = {Publisher: The American Astronomical Society},
	pages = {L38},
}

@article{rivera_situ_2024,
	title = {In situ observations of large-amplitude {Alfv{\'e}n} waves heating and accelerating the solar wind},
	volume = {385},
	issn = {0036-8075, 1095-9203},
	url = {https://www.science.org/doi/10.1126/science.adk6953},
	doi = {10.1126/science.adk6953},
	language = {en},
	number = {6712},
	urldate = {2024-09-02},
	journal = {Science},
	author = {Rivera, Yeimy J. and Badman, Samuel T. and Stevens, Michael L. and Verniero, Jaye L. and Stawarz, Julia E. and Shi, Chen and Raines, Jim M. and Paulson, Kristoff W. and Owen, Christopher J. and Niembro, Tatiana and Louarn, Philippe and Livi, Stefano A. and Lepri, Susan T. and Kasper, Justin C. and Horbury, Timothy S. and Halekas, Jasper S. and Dewey, Ryan M. and De Marco, Rossana and Bale, Stuart D.},
	month = aug,
	year = {2024},
	pages = {962--966},
}

@article{liu_determination_2021,
	title = {Determination of {Solar} {Wind} {Angular} {Momentum} and {Alfv{\'e}n} {Radius} from {Parker} {Solar} {Probe} {Observations}},
	volume = {908},
	issn = {2041-8205},
	url = {https://dx.doi.org/10.3847/2041-8213/abe38e},
	doi = {10.3847/2041-8213/abe38e},
	language = {en},
	number = {2},
	urldate = {2024-09-01},
	journal = {The Astrophysical Journal Letters},
	author = {Liu, Ying D. and Chen, Chong and Stevens, Michael L. and Liu, Mingzhe},
	month = feb,
	year = {2021},
	note = {Publisher: The American Astronomical Society},
	pages = {L41},
}

@article{liu_solar_2021,
	title = {Solar wind energy flux observations in the inner heliosphere: first results from {Parker} {Solar} {Probe}},
	volume = {650},
	copyright = {{\textcopyright} M. Liu et al. 2021},
	issn = {0004-6361, 1432-0746},
	shorttitle = {Solar wind energy flux observations in the inner heliosphere},
	url = {https://www.aanda.org/articles/aa/abs/2021/06/aa39615-20/aa39615-20.html},
	doi = {10.1051/0004-6361/202039615},
	language = {en},
	urldate = {2024-09-01},
	journal = {Astronomy \& Astrophysics},
	author = {Liu, M. and Issautier, K. and Meyer-Vernet, N. and Moncuquet, M. and Maksimovic, M. and Halekas, J. S. and Huang, J. and Griton, L. and Bale, S. and Bonnell, J. W. and Case, A. W. and Goetz, K. and Harvey, P. R. and Kasper, J. C. and MacDowall, R. J. and Malaspina, D. M. and Pulupa, M. and Stevens, M. L.},
	month = jun,
	year = {2021},
	note = {Publisher: EDP Sciences},
	pages = {A14},
}

@article{liu_total_2023,
	title = {Total electron temperature derived from quasi-thermal noise spectroscopy in the pristine solar wind from {Parker} {Solar} {Probe} observations},
	volume = {674},
	copyright = {{\textcopyright} The Authors 2023},
	issn = {0004-6361, 1432-0746},
	url = {https://www.aanda.org/articles/aa/abs/2023/06/aa45450-22/aa45450-22.html},
	doi = {10.1051/0004-6361/202245450},
	language = {en},
	urldate = {2024-09-01},
	journal = {Astronomy \& Astrophysics},
	author = {Liu, M. and Issautier, K. and Moncuquet, M. and Meyer-Vernet, N. and Maksimovic, M. and Huang, J. and Martinovic, M. M. and Griton, L. and Chrysaphi, N. and Jagarlamudi, V. K. and Bale, S. D. and Pulupa, M. and Kasper, J. C. and Stevens, M. L.},
	month = jun,
	year = {2023},
	note = {Publisher: EDP Sciences},
	pages = {A49},
}

@article{bruno_coordinated_2009,
	title = {Coordinated {Study} on {Solar} {Wind} {Turbulence} {During} the {Venus}-{Express}, {ACE} and {Ulysses} {Alignment} of {August} 2007},
	volume = {104},
	issn = {1573-0794},
	url = {https://doi.org/10.1007/s11038-008-9272-9},
	doi = {10.1007/s11038-008-9272-9},
	language = {en},
	number = {1},
	urldate = {2024-11-13},
	journal = {Earth, Moon, and Planets},
	author = {Bruno, R. and Carbone, V. and V{\"o}r{\"o}s, Z. and D{\textquoteright}Amicis, R. and Bavassano, B. and Cattaneo, M. B. and Mura, A. and Milillo, A. and Orsini, S. and Veltri, P. and Sorriso-Valvo, L. and Zhang, T. and Biernat, H. and Rucker, H. and Baumjohann, W. and Jankovi{\v c}ov{\'a}, D. and Kov{\'a}cs, P.},
	month = apr,
	year = {2009},
	pages = {101--104},
}

@article{huang_dominance_2024,
	title = {Dominance of 2 {Minute} {Oscillations} near the {Alfv{\'e}n} {Surface}},
	volume = {977},
	issn = {2041-8205},
	url = {https://dx.doi.org/10.3847/2041-8213/ad9271},
	doi = {10.3847/2041-8213/ad9271},
	language = {en},
	number = {1},
	urldate = {2024-12-05},
	journal = {The Astrophysical Journal Letters},
	author = {Huang, Zesen and Velli, Marco and Shi, Chen and Zhu, Yingjie and Chandran, B. D. G. and Bowen, Trevor and R{\'e}ville, Victor and Huang, Jia and Hou, Chuanpeng and Sioulas, Nikos and Liu, Mingzhe and Pulupa, Marc and Huang, Sheng and Bale, Stuart D.},
	month = dec,
	year = {2024},
	note = {Publisher: The American Astronomical Society},
	pages = {L12},
}

@article{didkovsky_observations_2011,
	title = {{OBSERVATIONS} {OF} {FIVE}-{MINUTE} {SOLAR} {OSCILLATIONS} {IN} {THE} {CORONA} {USING} {THE} {EXTREME} {ULTRAVIOLET} {SPECTROPHOTOMETER} ({ESP}) {ON} {BOARD} {THE} {SOLAR} {DYNAMICS} {OBSERVATORY} {EXTREME} {ULTRAVIOLET} {VARIABILITY} {EXPERIMENT} ({SDO}/{EVE})},
	volume = {738},
	issn = {2041-8205},
	url = {https://dx.doi.org/10.1088/2041-8205/738/1/L7},
	doi = {10.1088/2041-8205/738/1/L7},
	language = {en},
	number = {1},
	urldate = {2025-03-08},
	journal = {The Astrophysical Journal Letters},
	author = {Didkovsky, L. and Judge, D. and Kosovichev, A. G. and Wieman, S. and Woods, T.},
	month = aug,
	year = {2011},
	note = {Publisher: The American Astronomical Society},
	pages = {L7},
}

@article{musman_vertical_1970,
	title = {Vertical {Velocities} and {Horizontal} {Wave} {Propagation} in the {Solar} {Photosphere}},
	volume = {13},
	issn = {0038-0938},
	url = {https://ui.adsabs.harvard.edu/abs/1970SoPh...13..261M},
	doi = {10.1007/BF00153548},
	urldate = {2025-03-11},
	journal = {Solar Physics},
	author = {Musman, Steven and Rust, David M.},
	month = aug,
	year = {1970},
	note = {ADS Bibcode: 1970SoPh...13..261M},
	pages = {261--286},
}

@article{chashei_five-minute_1999,
	title = {Five-minute magnetic field fluctuations in the solar wind acceleration region},
	volume = {189},
	issn = {1573-093X},
	url = {https://doi.org/10.1023/A:1005223531849},
	doi = {10.1023/A:1005223531849},
	language = {en},
	number = {2},
	urldate = {2025-03-13},
	journal = {Solar Physics},
	author = {Chashei, I.V. and Bird, M.K. and Efimov, A.I. and Andreev, V.E. and Samoznaev, L.N.},
	month = nov,
	year = {1999},
	pages = {399--412},
}

@article{deubner_helioseismology_1984,
	title = {Helioseismology: {Oscillations} as a {Diagnostic} of the {Solar} {Interior}},
	volume = {22},
	issn = {0066-4146},
	shorttitle = {Helioseismology},
	url = {https://ui.adsabs.harvard.edu/abs/1984ARA&A..22..593D},
	doi = {10.1146/annurev.aa.22.090184.003113},
	urldate = {2025-03-16},
	journal = {Annual Review of Astronomy and Astrophysics},
	author = {Deubner, Franz-Ludwig and Gough, Douglas},
	month = jan,
	year = {1984},
	note = {ADS Bibcode: 1984ARA\&A..22..593D},
	pages = {593--619},
}

@article{bahcall_solar_1988,
	title = {Solar models, neutrino experiments, and helioseismology},
	volume = {60},
	url = {https://link.aps.org/doi/10.1103/RevModPhys.60.297},
	doi = {10.1103/RevModPhys.60.297},
	number = {2},
	urldate = {2025-03-16},
	journal = {Reviews of Modern Physics},
	author = {Bahcall, John N. and Ulrich, Roger K.},
	month = apr,
	year = {1988},
	note = {Publisher: American Physical Society},
	pages = {297--372},
}

@article{guglielmi_five-minute_2015,
	title = {Five-{Minute} {Solar} {Oscillations} and {Ion}-{Cyclotron} {Waves} in the {Solar} {Wind}},
	volume = {290},
	issn = {1573-093X},
	url = {https://doi.org/10.1007/s11207-015-0772-2},
	doi = {10.1007/s11207-015-0772-2},
	language = {en},
	number = {10},
	urldate = {2025-03-16},
	journal = {Solar Physics},
	author = {Guglielmi, A. and Potapov, A. and Dovbnya, B.},
	month = oct,
	year = {2015},
	pages = {3023--3032},
}

@article{calabro_oscillatory_2013,
	title = {Oscillatory {Behavior} in the {Corona}},
	volume = {286},
	issn = {1573-093X},
	url = {https://doi.org/10.1007/s11207-013-0283-y},
	doi = {10.1007/s11207-013-0283-y},
	language = {en},
	number = {2},
	urldate = {2025-03-16},
	journal = {Solar Physics},
	author = {Calabro, B. and McAteer, R. T. J. and Bloomfield, D. S.},
	month = sep,
	year = {2013},
	pages = {405--415},
}

@article{thomson_propagation_1995,
	title = {Propagation of solar oscillations through the interplanetary medium},
	volume = {376},
	copyright = {1995 Springer Nature Limited},
	issn = {1476-4687},
	url = {https://www.nature.com/articles/376139a0},
	doi = {10.1038/376139a0},
	language = {en},
	number = {6536},
	urldate = {2025-03-24},
	journal = {Nature},
	author = {Thomson, David J. and Maclennan, Carol G. and Lanzerotti, Louis J.},
	month = jul,
	year = {1995},
	note = {Publisher: Nature Publishing Group},
	pages = {139--144},
}

@misc{huang_two_2025,
	title = {Two {Types} of \$1/f\$ {Range} in {Solar} {Wind} {Turbulence}},
	url = {https://ui.adsabs.harvard.edu/abs/2025arXiv250617523H},
	doi = {10.48550/arXiv.2506.17523},
	urldate = {2025-06-26},
	publisher = {arXiv},
	author = {Huang, Zesen and Velli, Marco and Chandran, B. D. G. and Shi, Chen and Ding, Yuliang and Matteini, Lorenzo and Choi, Kyung-Eun},
	month = jun,
	year = {2025},
	note = {ADS Bibcode: 2025arXiv250617523H},
}

@article{morton_origins_2025-1,
	title = {On the {Origins} of {Coronal} {Alfv{\'e}nic} {Waves}},
	volume = {986},
	issn = {2041-8205},
	url = {https://dx.doi.org/10.3847/2041-8213/add7da},
	doi = {10.3847/2041-8213/add7da},
	language = {en},
	number = {1},
	urldate = {2025-07-11},
	journal = {The Astrophysical Journal Letters},
	author = {Morton, Richard J. and Soler, Roberto},
	month = jun,
	year = {2025},
	note = {Publisher: The American Astronomical Society},
	pages = {L6},
}

@article{pontieu_how_2005,
	title = {How to {Channel} {Photospheric} {Oscillations} into the {Corona}},
	volume = {624},
	issn = {0004-637X},
	url = {https://iopscience.iop.org/article/10.1086/430345/meta},
	doi = {10.1086/430345},
	language = {en},
	number = {1},
	urldate = {2025-08-12},
	journal = {The Astrophysical Journal},
	author = {Pontieu, B. De and Erd{\'e}lyi, R. and Moortel, I. De},
	month = apr,
	year = {2005},
	note = {Publisher: IOP Publishing},
	pages = {L61},
}

@article{miriyala_coronal_2025,
	title = {The {Coronal} {Power} {Spectrum} from {MHD} {Mode} {Conversion} above {Sunspots}},
	volume = {979},
	issn = {0004-637X, 1538-4357},
	url = {https://iopscience.iop.org/article/10.3847/1538-4357/ada26f},
	doi = {10.3847/1538-4357/ada26f},
	number = {2},
	urldate = {2025-08-14},
	journal = {The Astrophysical Journal},
	author = {Miriyala, Hemanthi and Morton, Richard J. and Khomenko, Elena and Antolin, Patrick and Botha, Gert J.J.},
	month = feb,
	year = {2025},
	pages = {236},
}
\bibliographystyle{sciencemag}

%
%
%
%
%
%


\section*{Acknowledgments}
Parker Solar Probe was designed, built, and is now operated by the Johns Hopkins Applied Physics Laboratory as part of NASA’s Living with a Star (LWS) program (contract NNN06AA01C). The SWEAP Investigation is a multi-institution project led by the Smithsonian Astrophysical Observatory in Cambridge, Massachusetts. Other members of the SWEAP team come from the University of Michigan, University of California, Berkeley Space Sciences Laboratory, The NASA Marshall Space Flight Center, The University of Alabama Huntsville, the Massachusetts Institute of Technology, Los Alamos National Laboratory, Draper Laboratory, JHU’s Applied Physics Laboratory, and NASA Goddard Space Flight Center. The FIELDS instrument suite was designed and built and is operated by a consortium of institutions including the University of California, Berkeley; University of Minnesota; University of Colorado, Boulder; NASA/GSFC, CNRS/LPC2E, University of New Hampshire, University of Maryland, UCLA, IFRU, Observatoire de Meudon, Imperial College, London, and Queen Mary University London. Z.H. acknowledges helpful discussion with Dr. Yingjie Zhu. Z.H. also acknowledges UCLA OpenAI project for providing access to ChatGPT. 
\paragraph*{Funding:}
This research was funded in part by the FIELDS experiment on the Parker Solar Probe spacecraft, designed and developed under NASA contract UCB \#00010350/NASA NNN06AA01C, and the NASA Parker Solar Probe Observatory Scientist grant NASA NNX15AF34G, and NASA HTMS 80NSSC20K1275. Z.H., M.V., and O.P. acknowledge NASA AIAH 80NSSC25K0386. B.C. and K.G.K. acknowledge support from NASA grant 80NSSC24K0171. B.C. acknowledges NASA grants NNN06AA01C. C.S. acknowledges supported from NSF SHINE \#2229566 and NASA ECIP \#80NSSC23K1064. Y.J.R., M.S., S.T.B., F.F. is partially supported by the Parker Solar Probe project through the SAO/SWEAP subcontract 975569. M.L. acknowledges support from NASA HGI-O grant \#80NSSC25K7689. J.H. thank the support by NASA LWS grant 80NSSC23K0737. T.E. acknowledges funding from The Chuck Lorre Family Foundation Big Bang Theory Graduate Fellowship and NASA grant 80NSSC20K1285. R.J.M. is supported by a UKRI Future Leader Fellowship (RiPSAW-MR/T019891/1). 

\paragraph*{Author contributions:}
Z.H. was responsible for the conceptualization, formal analysis and writing the manuscript. M.V. helped conceive and design the study. O.P. help analyzed the magnetic connectivity. O.P., R.J.M., C.S., Y.R., B.C., Y.D., T.E., C.H., S.T.B. F.F. contributed interpretation of the spacecraft data. N.R., S.T.B., M.S. S.D.B. K.G.K., O.R., J.H., M.L., D.E.L., M.P., R.L. performed data acquisition. All authors commented on the manuscript.
\paragraph*{Competing interests:}
There are no competing interests to declare.
\paragraph*{Data and materials availability:}
For Parker Solar Probe the data is available at https://spdf.gsfc.nasa.gov/pub/data/psp/. The orbital data is available at https://psp-gateway.jhuapl.edu/.


\subsection*{Supplementary materials}
Materials and Methods\\
Figs. S1 to S6\\
References \textit{(7-\arabic{enumiv})}\\ 


\newpage


\renewcommand{\thefigure}{S\arabic{figure}}
\renewcommand{\thetable}{S\arabic{table}}
\renewcommand{\theequation}{S\arabic{equation}}
\renewcommand{\thepage}{S\arabic{page}}
\setcounter{figure}{0}
\setcounter{table}{0}
\setcounter{equation}{0}
\setcounter{page}{1} 


\begin{center}
\section*{Supplementary Materials for\\ \scititle}

Zesen~Huang$^{\ast\dagger}$\\ 
\small$^\ast$Corresponding author. Email: zesenhuang@g.ucla.edu\\
\small$^\dagger$These authors contributed equally to this work.
\end{center}

\subsubsection*{This PDF file includes:}
Materials and Methods\\
Figures S1 to S6\\


\newpage


\subsection*{Materials and Methods}


\subsection*{Supplementary Text}

\subsubsection*{Coordinate System}
For Parker Solar Probe we adopt the heliocentric Radial--Tangential--Normal (RTN) frame, in which the radial unit vector points from the Sun to the spacecraft, 
$\hat{\mathbf{R}}=\mathbf{r}/\|\mathbf{r}\|$; the tangential unit vector is set by the Sun’s rotation axis (unit vector $\hat{\boldsymbol{\Omega}}_\odot$) via 
$\hat{\mathbf{T}}=\frac{\hat{\boldsymbol{\Omega}}_\odot \times \hat{\mathbf{R}}}{\|\hat{\boldsymbol{\Omega}}_\odot \times \hat{\mathbf{R}}\|}$ 
(lying in the solar equatorial plane and positive in the corotational/solar-west direction); and the normal completes the right-handed triad, 
$\hat{\mathbf{N}}=\hat{\mathbf{R}}\times\hat{\mathbf{T}}$. 
This definition does not assume circular motion and is the standard used in PSP data products for reporting plasma and magnetic-field components $(V_r,V_t,V_n)$ and $(B_r,B_t,B_n)$.

\subsubsection*{PSP \textit{in situ} Measurements}
We use magnetic field data from the fluxgate magnetometer in the FIELDS instrument suite \cite{bale_fields_2016,bowen_merged_2020} and plasma moments from SPAN-i and SPAN-a in the SWEAP instrument suite \cite{kasper_solar_2016, livi_solar_2022}. The radial component of the first-order proton (partial) moment is treated as the solar wind speed. Electron number density data from Quasi-Thermal-Noise (QTN) spectroscopy serve as a critical calibration standard \cite{kasper_parker_2021,zhao_analysis_2021,liu_solar_2021,liu_determination_2021,liu_total_2023,zheng_solar_2024}, as alpha-particle abundance is typically negligible (2\%–6\%) \cite{mostafavi_alphaproton_2022}. Thus, we use QTN-derived electron number density as a proxy for proton number density when available. Otherwise, the proton number density product in SPAN $n_{p,SPAN}$ is used.

\begin{figure}
	\centering
	\includegraphics[width=1.0\textwidth]{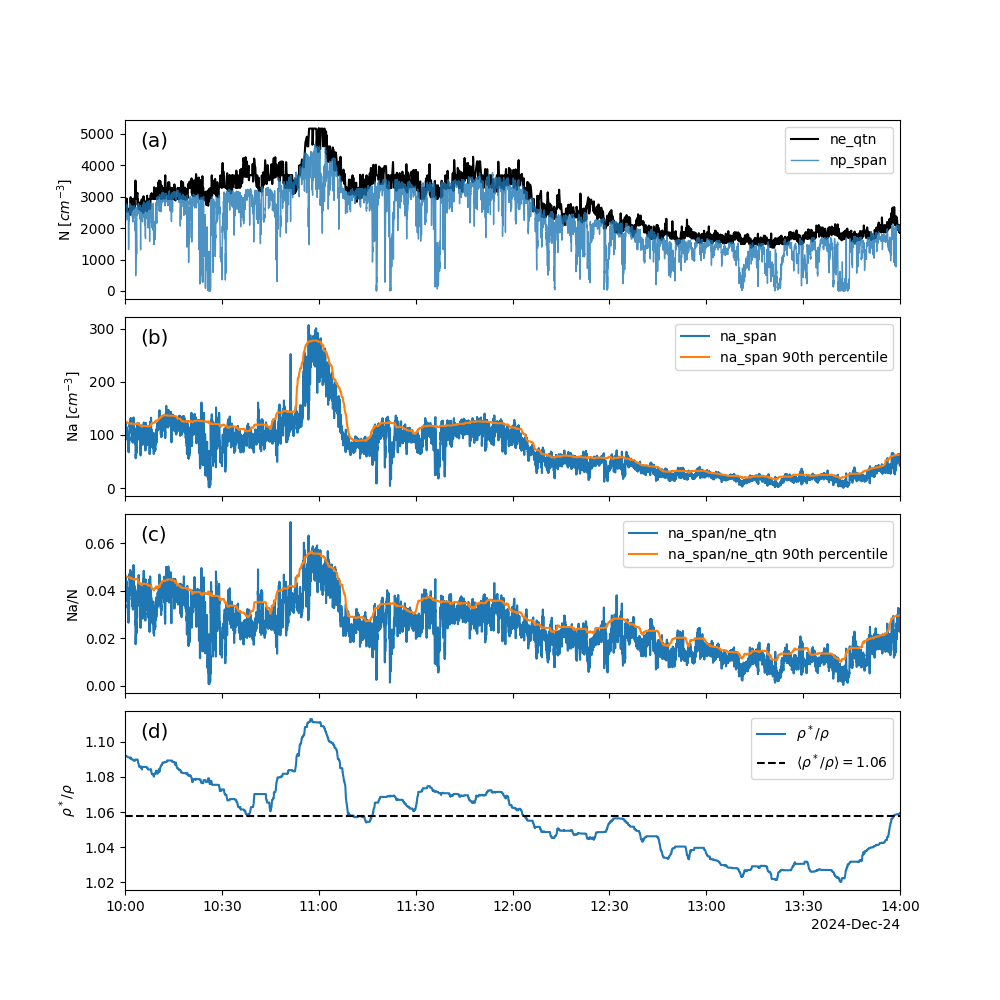}
	\caption{\textbf{Comparison between QTN and SPAN-i} \textbf{(a)} Electron number density (ne-qtn) from Quasi-Thermal-Noise (QTN) and proton number density from SPAN-i (np-span). \textbf{(b)} Alpha particle number density (na-span) from SPAN-i. \textbf{(c)} Ratio between ne-qtn and na-span. \textbf{(d)} Ratio between the mass density modified with alpha particle density ($\rho^*$) and without modification ($\rho$).}
	\label{fig:na_np_ratio}
\end{figure}

\begin{figure} 
	\centering
	\includegraphics[width=1.0\textwidth]{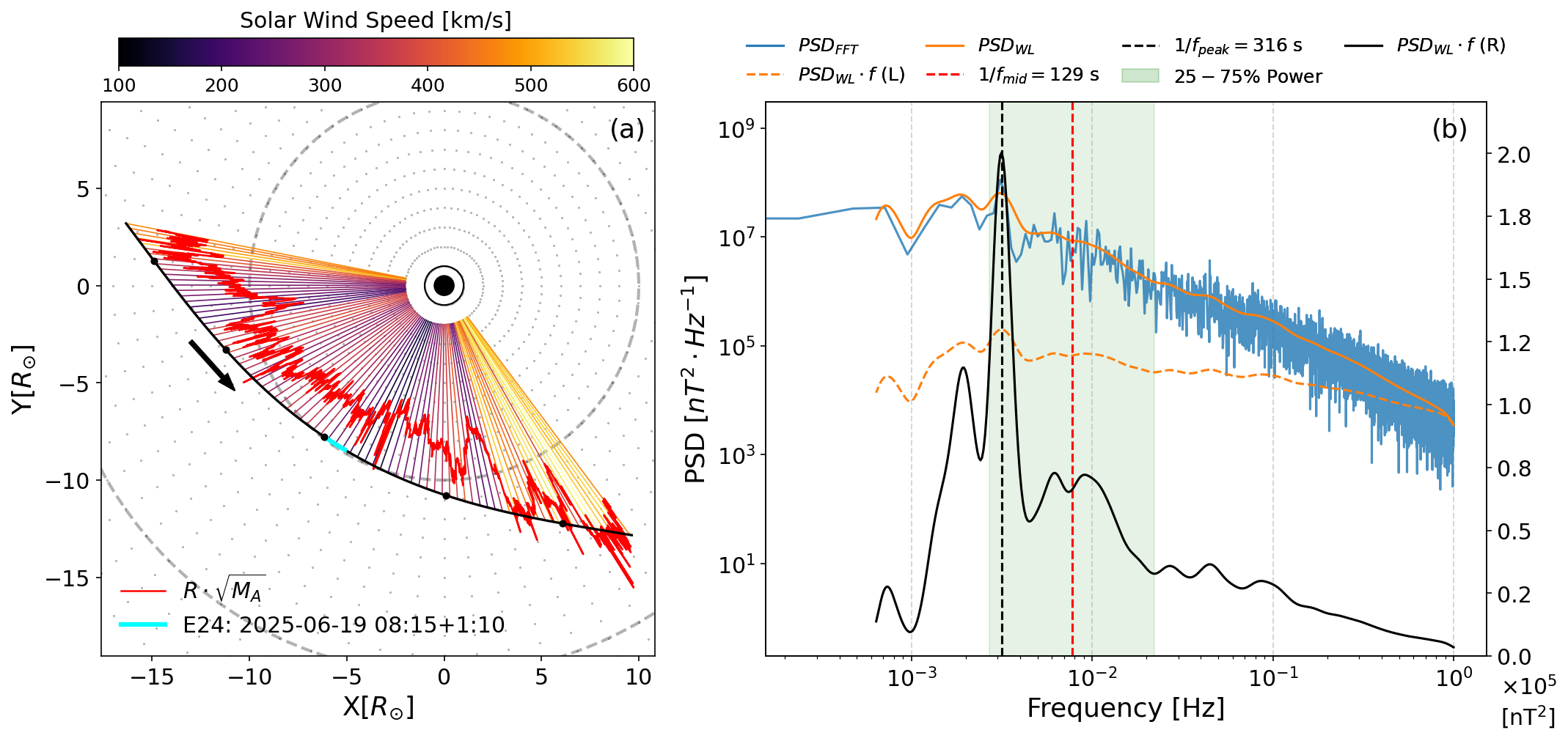} 

	\caption{\textbf{5-Minute Oscillation Event from PSP E24}
		Similar to Figure~\ref{fig:event-p1}, but for E24 event. Cyan bar marks the selected interval (2025-06-19 08:15–09:25).}
	\label{fig:sup_E24_event-p1} 
\end{figure}

\begin{figure} 
	\centering
	\includegraphics[width=1.0\textwidth]{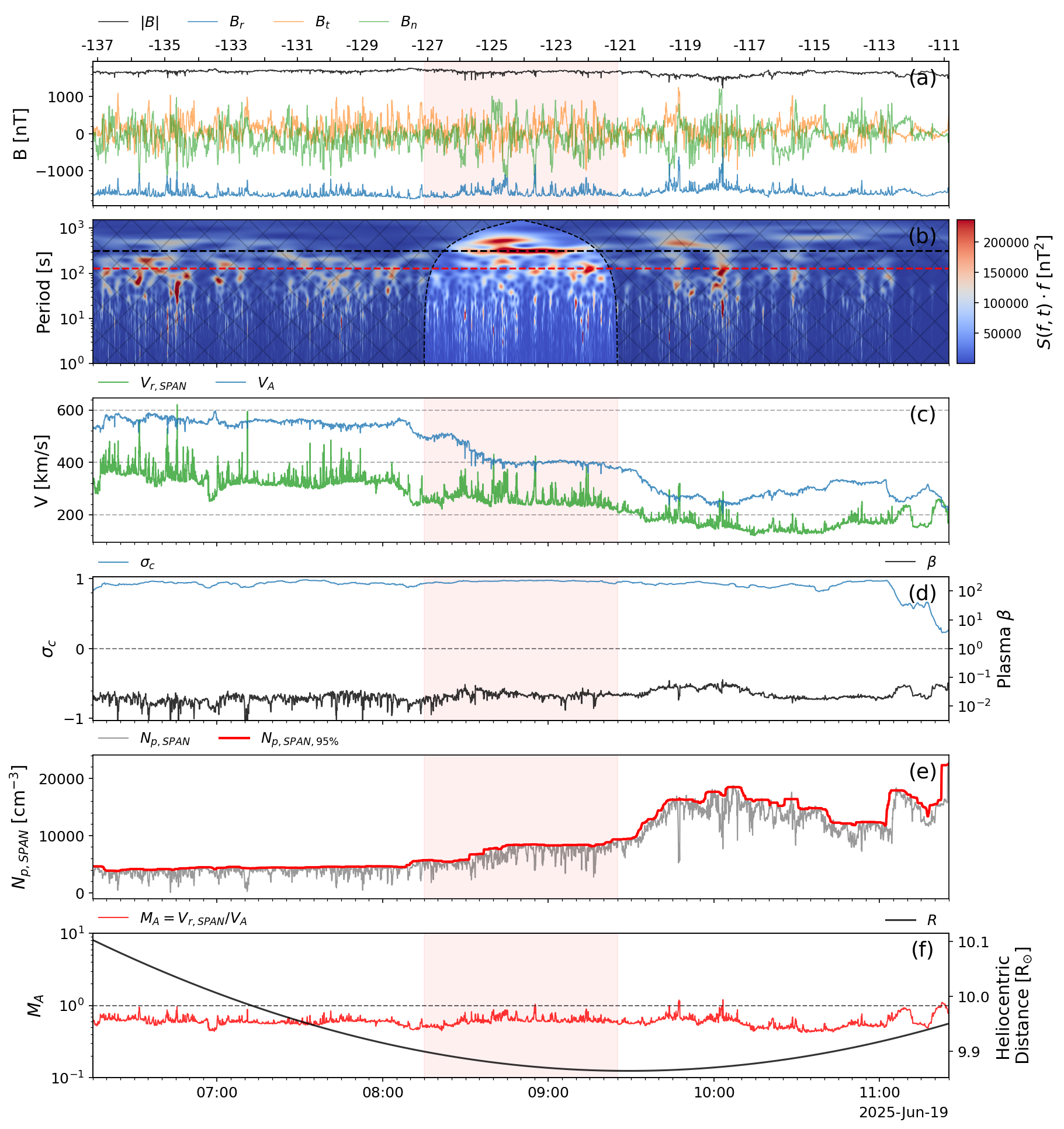} 

	\caption{\textbf{Timeseries of E24 Event}
		Similar to Figure~\ref{fig:event-p2}, but for E24 event.}
	\label{fig:sup_E24_event-p2} 
\end{figure}

\begin{figure} 
	\centering
	\includegraphics[width=1.0\textwidth]{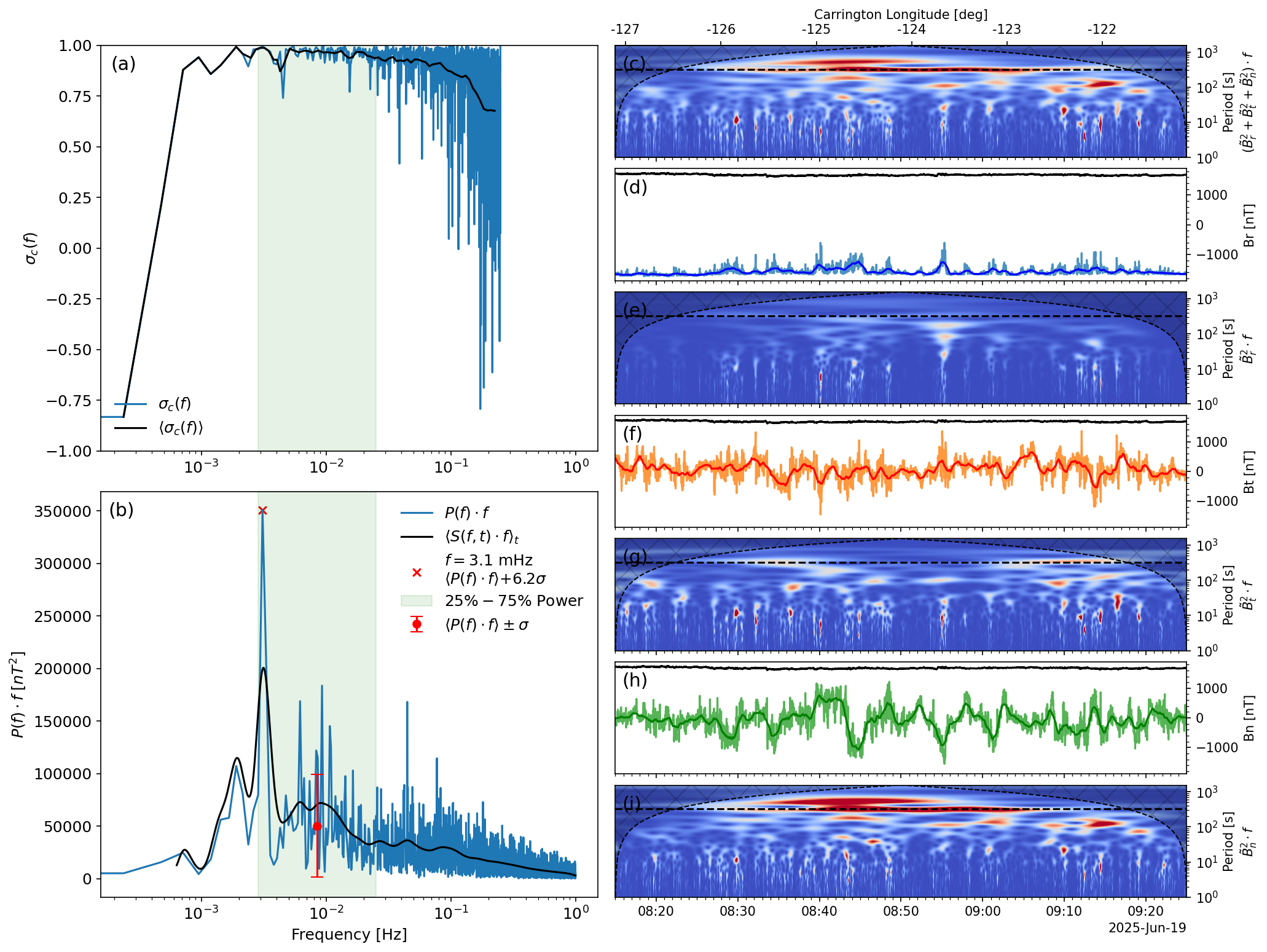} 
	\caption{Similar to Fig. \ref{fig:sigma_c} but for E24 event.}
	\label{fig:sup_E24_polarization} 
\end{figure}

Figure \ref{fig:na_np_ratio} compares electron number density from QTN with proton and alpha particle number densities from SPAN-i. Due to the limited field of view of SPAN-i, their partial moments often underestimate particle densities, as seen in the np-span dropouts in \textbf{(a)}. However, the upper envelope of np-span aligns well with ne-qtn, confirming that SPAN-i rarely overestimates proton density.

For simplicity, we use electron number density as a proxy for proton number density, though this underestimates mass density due to alpha particles. The alpha particle number density, shown in \textbf{(b)}, is also underestimated. Using the upper envelope (1-minute window) of na-span, we estimate the ratio $N_a/N_e$, shown in \textbf{(c)}. Assuming charge neutrality ($N_e = 2N_a + N_p$), the mass density $\rho$ is calculated as:
\begin{eqnarray}
	\rho = N_e \cdot m_p
\end{eqnarray}
A more accurate mass density $\rho^*$, accounting for alpha particles, is:
\begin{eqnarray}
	\rho^* = N_p \cdot m_p + 4 \cdot N_a \cdot m_p
\end{eqnarray}
The ratio $\rho^*/\rho$ is:
\begin{eqnarray}
	\rho^*/\rho = 1 + 2\frac{N_a}{N_e}
\end{eqnarray}
This ratio, shown in \textbf{(d)}, indicates that $\rho$ is underestimated by $\sim$6\% (or $\sim$8\% for 10:00–12:00). Since $\rho$ is only used to calculate cross helicity, where a 10\% change in mass density has minimal impact, we use $\rho$ throughout for simplicity and clarity. Throughout the text, we use $N_{p,QTN}$ to refer to $N_e$ derived via QTN technique. This quantity is only used for the calculation of the cross helicity and plasma $\beta$. In this study, both quantities are not sensitive to the data product being used.

\subsubsection*{Cross Helicity}
In this work, we use cross helicity $\sigma_c$ to estimate the Alfv\'enicity of the solar wind timeseries. $\sigma_c$ is calculated in two forms: $\sigma_c(t)$, which shows the temporal correlation between \textbf{$\vec{B}$} and \textbf{$\vec{V}$}, and $\sigma_c(f)$, which visualizes their frequency-dependent correlation.

To compute $\sigma_c(t)$ (e.g., Fig. \ref{fig:event-p2}\textbf{(d)}), we first convert the magnetic field into velocity units:
\begin{eqnarray}
	\boldsymbol{Va} = \boldsymbol{B}/\sqrt{\mu_0 \rho},
\end{eqnarray}
where $\rho = N_{e,QTN} \cdot m_p$. The fluctuating components are obtained by subtracting the 10-minute rolling average:
\begin{eqnarray}
	\delta Va_{r,t,n} = Va_{r,t,n} - \langle Va_{r,t,n} \rangle_{10\mathrm{min}}, \\
	\delta V_{r,t,n} = V_{r,t,n} - \langle V_{r,t,n} \rangle_{10\mathrm{min}},
\end{eqnarray}
where $V$ is the proton velocity from SPAN-i. The Els\"asser variables are then defined as:
\begin{eqnarray}
	\boldsymbol{z}^{\pm}(t) = \delta \boldsymbol{V} \mp \delta \boldsymbol{Va}.
\end{eqnarray}
Finally, the cross helicity is:
\begin{eqnarray}
	\sigma_c(t) = \frac{{z^+}^2 - {z^-}^2}{{z^+}^2 + {z^-}^2}.
\end{eqnarray}

To compute $\sigma_c(f)$ (e.g., Fig. \ref{fig:sigma_c}\textbf{(a)}), we take the Fast Fourier transform of the velocity and magnetic field using NumPy \cite{harris_array_2020}:
\begin{eqnarray}
	\widehat{\boldsymbol{Va}}(f) = \mathcal{F}(\boldsymbol{Va}), \\
	\widehat{\boldsymbol{V}}(f) = \mathcal{F}(\boldsymbol{V}).
\end{eqnarray}
The frequency-domain Els\"asser variables are:
\begin{eqnarray}
	\boldsymbol{z}^{\pm}(f) = \widehat{\boldsymbol{V}} \mp \widehat{\boldsymbol{Va}},
\end{eqnarray}
and the cross helicity is:
\begin{eqnarray}
	\sigma_c(f) = \frac{{z^+}^2 - {z^-}^2}{{z^+}^2 + {z^-}^2}.
\end{eqnarray}

In summary, $\sigma_c(t)$ captures the correlation between $\boldsymbol{V}$ and $\boldsymbol{B}$ for fluctuations with periods shorter than 10 minutes, while $\sigma_c(f)$ provides a frequency-resolved view of this correlation.

\subsubsection*{Alfv\'en Speed}
The Alfv\'en speed is defined as follows:
\begin{eqnarray}
    V_A = |B|/\sqrt{\mu_0 m_p n_p}
\end{eqnarray}
where $\mu_0$ is the vacuum permeability, $m_p$ is proton mass and $n_p$ is the proton number density. When $n_{e,QTN}$ is available, it is used as the value for $n_p$ (E22 case). Otherwise, a 1-minute rolling window 90\% quantile will be applied to $n_{p,SPAN}$. This is to remove the instrument related field of view issue. The result is used for $n_p$ (E24 case). The Alfv\'en speed in this study is only used for quantifying the Alfv\'en mach number. The primary results are not related to the actual values of $V_A$.

\subsubsection*{Plasma $\beta$}
The (proton) plasma $\beta$ is defined as follows:
\begin{eqnarray}
	\beta = \frac{n_p k_B T_p}{B^2/2\mu_0}
\end{eqnarray}
where $n_p$ is the proton number density, $k_B$ is the Boltzmann constant, $T_p$ is the proton temperature, $B$ is magnetic field magnitude, $\mu_0$ is the vacuum permeability. Here we use $n_{e,QTN}$ as a proxy for $n_p$ when available, otherwise $n_{p,SPAN}$ is used. 

\subsubsection*{Power Spectral Density}
In this work, we use two methods to calculate the power spectral density (PSD) of the magnetic field. The magnetic field vectors are first resampled to 500 ms. 

The first method is the Fast Fourier Transform (FFT) using NumPy \cite{harris_array_2020}:
\begin{eqnarray}
	\widehat{B}_{r,t,n} = \mathcal{F}(B_{r,t,n}),
\end{eqnarray} 
where the trace magnetic PSD is computed as:
\begin{eqnarray}
	\mathrm{PSD}_{FFT} = \widehat{B}_r^2 + \widehat{B}_t^2 + \widehat{B}_n^2.
\end{eqnarray}

The second method is the continuous wavelet transform (CWT) using PyCWT. Each component of $\boldsymbol{B}$ is transformed into $S_{r,t,n}(f,t)$ using the default Morlet wavelet ($f_0=6$). The time-averaged wavelet PSD is then calculated as:
\begin{eqnarray}
	\mathrm{PSD}_{WL} = \langle S_{r}(f,t)^2 + S_{t}(f,t)^2 + S_{n}(f,t)^2 \rangle_t,
\end{eqnarray}
which serves as a smoothed version of $\mathrm{PSD}_{FFT}$ (see, e.g., Fig. \ref{fig:event-p1}). The wavelet transform using a Morlet wavelet is nearly equivalent to a windowed FFT and conserves total energy.

To enable a fair comparison between different frequencies on a logarithmic scale, we rectify the spectrum with $f$ and plot it in a linear-log scale. This is because:
\begin{eqnarray}
	P(f)\ \mathrm{d}f = P(f) f \ \mathrm{d} \ln f.
\end{eqnarray}
The area under $P(f)\cdot f$ in the linear-log axis represents the distribution of fluctuation energy across logarithmic frequency intervals.

\subsubsection*{Magnetic Connectivity}
Magnetic field connectivity with the solar sources during E22 and E24 perihelia. The coronal magnetic field is reconstructed using the Potential Field Source Surface Modeling (PFSS) \cite{panasenco_exploring_2020}. The thick black lines are the model neutral lines. Black contours indicate magnetic field pressure at 1.1 $R_s$ ($R_{\odot}$). The ballistic projection of the PSP trajectory (blue diamonds) on the source surface (blue crosses) and down to the solar wind source regions (blue circles) is calculated for source surfaces $R_{ss}/R_s$ = 2.5 (see \cite{panasenco_exploring_2020} for details) and measured in situ solar wind speed $\pm$ 80 $km/s$. Open magnetic field regions are shown in blue (negative) and green (positive). For both cases, solar wind was magnetically connected to a low-latitude coronal hole in the southern solar hemisphere.

\begin{figure}
	\centering
	\includegraphics[width = 0.7 \textwidth]{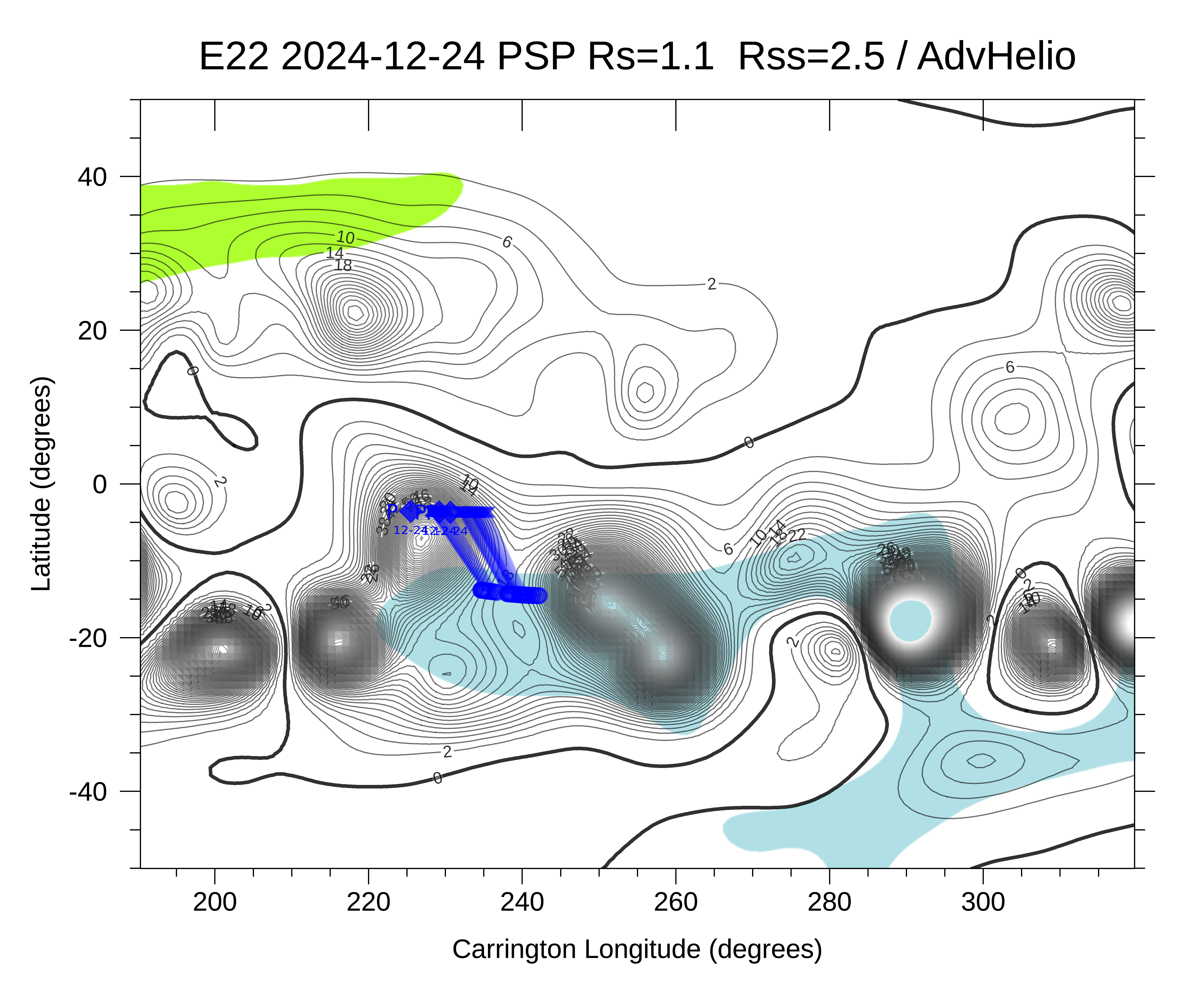}
	\caption{
		PFSS model of E22.
	}
	\label{fig:pfss_e22}
\end{figure}

\begin{figure}
	\centering
	\includegraphics[width=0.7\textwidth]{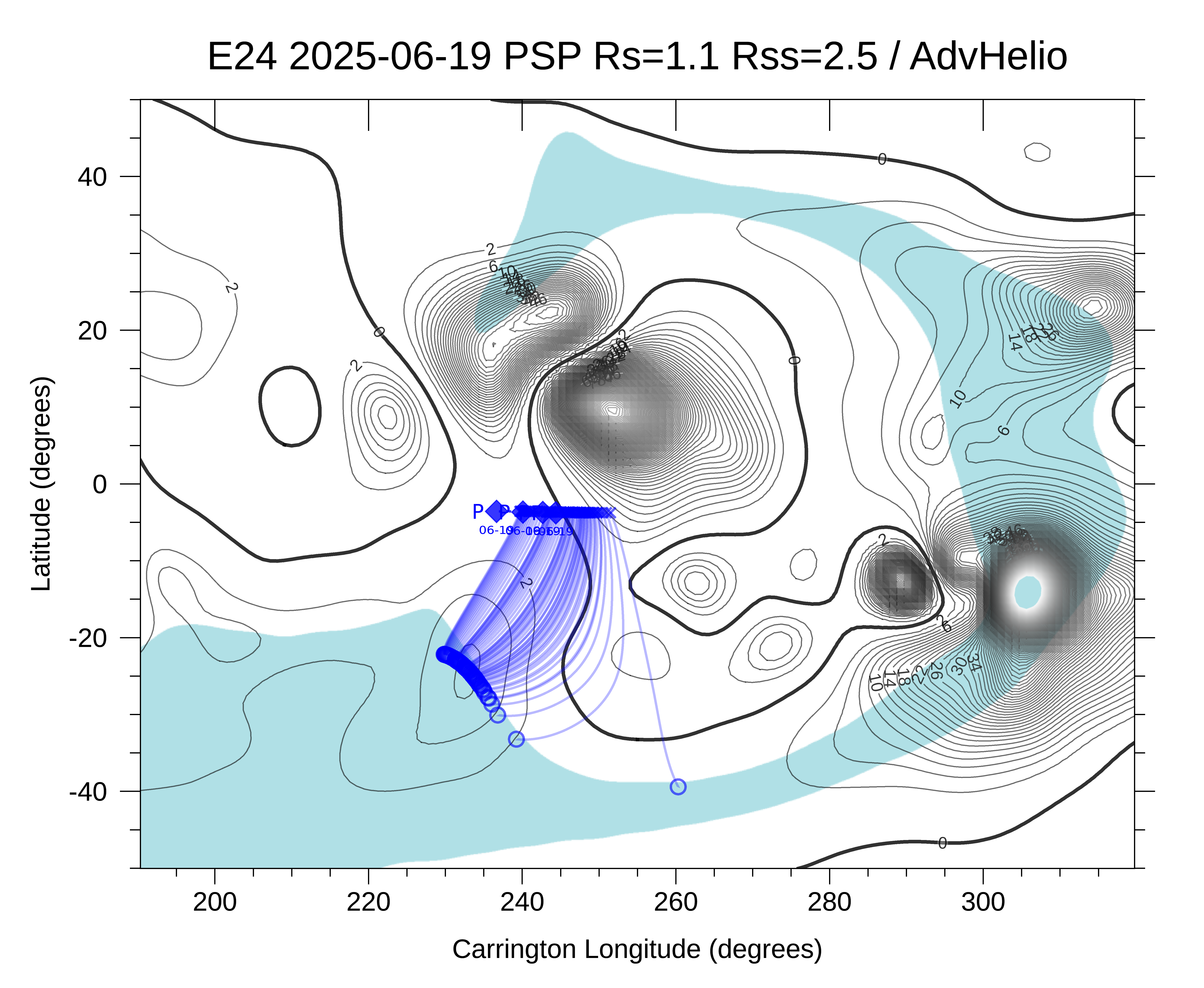}
	\caption{PFSS model of E24.}
	\label{fig:pfss_e24}
\end{figure}

\end{document}